\documentclass[twocolumn,aps,prx,amsmath,amssymb,superscriptaddress,groupedaddress]{revtex4-2}

\usepackage{graphicx}
\usepackage{dcolumn}
\usepackage{bm}
\usepackage{color}
\usepackage{textcomp}
\usepackage{amsthm}

\def\di{{\rm d}}
\def\eu{{\rm e}}
\def\iu{{\rm i}}

\usepackage{microtype}
\usepackage[hyperindex=true,pdftitle={The Origin of Dynamical Quantum Non-locality},
colorlinks=true,pagebackref=false,citecolor=blue,plainpages=false,pdfpagelabels,breaklinks=true,
linkcolor=blue,urlcolor=blue]{hyperref}

\newtheorem{theorem}{Theorem}
\newtheorem{lemma}{Lemma}
\newtheorem{corollary}{Corollary}

\begin{document}

\title{The Origin of Dynamical Quantum Non-locality}

\author{C\'esar E. Pach\'on}
\affiliation{guane Enterprises, R+D+I Unit, Medell\'in 050010, Colombia}
\author{Leonardo A. Pach\'on}
\affiliation{guane Enterprises, R+D+I Unit, Medell\'in 050010, Colombia}

\begin{abstract}
Non-locality is one of the hallmarks of quantum mechanics and is responsible for paradigmatic features such as entanglement and the Aharonov-Bohm effect. Non-locality comes in two flavours: a \emph{kinematic} non-locality---arising from the structure of the Hilbert space---and a \emph{dynamical} non-locality---arising from the quantum equations of motion. Recently, the origin of kinematic non-locality was traced to the uncertainty principle; here we rigorously trace the origin of dynamical non-locality to the superposition principle. We prove, via deformation quantization and Marinov's phase-space path integrals, that the exact Wigner propagator reduces to the classical Liouville propagator if and only if the Hamiltonian has at-most-quadratic Weyl symbol. This unified theorem covers both continuous-variable and finite-dimensional Hilbert spaces, aligning the Gaussian (CV) and Clifford (finite-$d$) boundaries of classical simulability into a single algebraic criterion. We introduce a macroscopic, experimentally accessible measure of dynamical non-locality---the signed divergence $\mathcal{D}(t)$---and show that it governs five phenomena: (i) the dynamical penalty incurred by quantum non-local games under post-measurement evolution; (ii) the quantum corrections to out-of-time-order correlators; (iii) the metrological gain beyond the shot-noise limit; (iv) the generation of non-Gaussian entanglement from product states; and (v) the non-Clifford / magic-state content of finite-dimensional dynamics. A concrete experimental protocol in circuit QED is proposed and complemented by a three-qubit CCZ protocol accessible on current qubit platforms.
\end{abstract}

\date{\today}
\maketitle

\section{Introduction}
\label{sec:intro}

Non-locality is one of the most intriguing features of quantum theory and comes in two distinct flavours. \emph{Kinematic} non-locality---the Hilbert-space content underlying Bell-type bounds~\cite{Bel66,AGR82,BC&10}---was traced to the fine-grained uncertainty principle by Oppenheim and Wehner~\cite{OW10}. \emph{Dynamical} non-locality---the non-local character of quantum equations of motion---was first identified by Aharonov and Bohm~\cite{AB59} and subsequently analysed through modular variables in the AB setting~\cite{APP69,Pop10,Tol11} and tested experimentally~\cite{OM&86}. Whereas the kinematic flavour has a well-established principle-level origin, the symmetric question for the dynamical flavour has remained open: its fundamental origin has often been conflated with the peculiarities of the displacement operator, the modular-variable framework has been largely confined to topological gauge phenomena, and semiclassical treatments of the Wigner propagator~\cite{DVS06,DGP10,Pac10,DP09} have not been distilled into a macroscopic, experimentally accessible observable.

Three questions therefore remain. Is dynamical non-locality a fundamental feature of quantum dynamics or merely an effective curiosity attached to the AB setting? What physical principle governs it? And what single observable can certify it on a concrete preparation?

In this work we answer all three. We prove that dynamical non-locality originates exclusively from the superposition principle, and that a single algebraic criterion on the Weyl symbol (at most quadratic) simultaneously characterises the Gaussian (continuous-variable) and Clifford (finite-dimensional) boundaries of classical simulability. We introduce a signed scalar observable---the divergence $\mathcal{D}(t)$---that certifies the violation of this criterion from a single survival-probability measurement, and demonstrate that it controls five apparently disparate phenomena: dynamical degradation of non-local games (Sec.~\ref{sec:games}); quantum corrections to out-of-time-order correlators (Sec.~\ref{sec:scrambling}); Heisenberg-scaling metrology (Sec.~\ref{sec:metrology}); generation of non-Gaussian entanglement from product states (Sec.~\ref{sec:entanglement}); and---in the discrete setting---the non-Clifford/magic-state content that underlies quantum computational advantage (Sec.~\ref{sec:clifford}). A concrete experimental protocol in circuit QED is proposed (Sec.~\ref{sec:cQED}), complemented by a parameter-free three-qubit CCZ signal with rational coefficient $c_{3}=1/64$. Throughout we treat unitary dynamics; decoherence is deferred to future work.

\section{The Origin of Dynamical Non-Locality}
\label{sec:origin}

To separate kinematic from dynamical effects cleanly, we employ the phase-space formulation of quantum mechanics (Supplementary Material~S0). The non-local character of quantum dynamics is fully encoded in the exact Wigner propagator $G_{\mathrm{W}}(\mathbf{r}'',t;\mathbf{r}',0)$. Its privileged role among dynamical objects stems from the fact that the underlying Weyl displacement operators generate the complete algebra of quantum observables~\cite{Wey27,Str05}; this explains why the dynamical non-locality first identified by Aharonov and Bohm~\cite{AB59} in modular variables is not an isolated curiosity of the displacement operator but a generic feature of the full algebra. In close analogy to Feynman's formulation~\cite{FH65}, the exact propagator admits the Marinov phase-space path-integral representation~\cite{Mar91,Pac10}
\begin{equation}
\label{equ:PathIntegral}
G_{\mathrm{W}}(\mathbf{r}'',\mathbf{r}')
= \frac{1}{h}\int {\cal D}^{2}r\,{\cal D}^{2}{\tilde r}\,
\exp\!\left(\frac{\iu}{\hbar} S[\{\mathbf{r}\},\{\tilde{\mathbf{r}}\},t]\right)\!,
\end{equation}
where the integration is over pairs of phase-space paths with $\mathbf{r}(0)=\mathbf{r}'$. For $\hat{H}=\hat{p}^{2}/2m+V(\hat{q})$, the action reads
\begin{equation}
\label{equ:Action}
S = \int_{0}^{t}\!\di t'\!\left[V\!\left(q + \tfrac{\tilde{q}}{2}\right) - V\!\left(q - \tfrac{\tilde{q}}{2}\right) - \tilde{q}\,\frac{\di V}{\di q}\right]\!.
\end{equation}
The variable $\tilde{q}$ parametrises the separation between paired paths; the superposition principle operates directly over this manifold. An analogous discrete path-separation variable $\xi$ appears in the Marinov integral on $\mathbb{Z}_{d}^{2}$ constructed in the companion paper~\cite{GP26}.

\emph{Theorem (Superposition origin of dynamical non-locality).}---\emph{In both continuous-variable phase space ($\mathbb{R}^{2f}$, with Weyl symbol $H$ real-analytic and satisfying $\sup_{\gamma}|\partial_{\gamma}^{k}H|/k!\le CR^{-k}$) and finite-dimensional phase space ($\mathbb{Z}_{d}^{2f}$, $d$ odd prime), the following are equivalent:} (i) \emph{$H$ is at most quadratic in the coordinates of $\Gamma$;} (ii) \emph{the Marinov action vanishes identically as a functional of the path-separation variable;} (iii) \emph{$G_{\mathrm{W}}$ coincides with the symplectic classical flow, $G_{\mathrm{W}}(\gamma'',t;\gamma',0)=\delta[\gamma''-\gamma^{\mathrm{cl}}(\gamma',t)]$.} Equivalently, $\Delta G_{\mathrm{W}}^{\mathrm{NL}}\neq 0$ iff $H$ carries cubic-or-higher content. In the CV case with standard kinetic term, condition~(i) reduces to $V$ at most quadratic; in the finite-$d$ case it is equivalent to $\hat{H}$ being a Clifford Hamiltonian on the generalised Pauli group~\cite{Gross2006,Gottesman1998}.

\emph{Proof idea (full statement and proof: Supplementary Material~S1).}---Both cases reduce to a single algebraic identity. Taylor-expanding $H(\gamma\pm\xi)$ in $\xi$,
\begin{equation}
\label{equ:TaylorKey}
H(\gamma+\xi)-H(\gamma-\xi) = 2\,\xi\cdot\nabla_{\gamma}H + \sum_{k\ge 1}\frac{2\,\xi^{\otimes(2k+1)}}{(2k+1)!}\cdot\partial^{2k+1}_{\gamma}H,
\end{equation}
the remainder sum vanishes identically in $\gamma$ and $\xi$ iff $\partial_{\gamma}^{3}H\equiv 0$, which (by analyticity in the CV case, by polynomial reduction over $\mathbb{Z}_{d}$ in the finite-$d$ case) forces $H$ at most quadratic. (i)$\Rightarrow$(ii): a quadratic $H$ kills all $k\ge 1$ terms of Eq.~(\ref{equ:TaylorKey}), so the bracket in the Marinov action~(\ref{equ:Action}) vanishes identically and $S\equiv 0$ as a functional of $\xi$. (ii)$\Rightarrow$(iii): a vanishing action reduces the path integral to $\int\mathcal{D}\xi\cdot 1$, enforcing the classical Hamilton equations and producing the symplectic $\delta$-function. (iii)$\Rightarrow$(i): if $\partial_{\gamma}^{3}H\not\equiv 0$, the leading cubic contribution $S\propto\xi^{3}$ in Eq.~(\ref{equ:TaylorKey}) generates Airy-type oscillatory corrections that broaden $G_{\mathrm{W}}$ away from the classical $\delta$-function.

As a concrete check: for $V=\lambda q^{3}$, Eq.~(\ref{equ:TaylorKey}) gives $S=\lambda\int_{0}^{t}\di t'\,\tilde{q}^{3}/4$, manifestly non-zero for $\tilde{q}\neq 0$. For the Kerr $V\propto\hat{q}^{4}$ analysed below, the leading contribution is $S\propto\tilde{q}^{3}\,q$. On a single qubit, every function on $\mathbb{Z}_{2}^{2}$ admits a multilinear polynomial representative of degree $\le 2$, so every single-qubit Hamiltonian is degenerately Clifford and $\mathcal{D}(t)\equiv 0$. Cubic Weyl content first arises on three qubits (Sec.~\ref{sec:clifford}).

The theorem establishes a sharp dichotomy. The uncertainty principle governs kinematic non-locality (the non-classical features of $\rho_{\mathrm{W}}$, including its negativity and sub-Planck structure), while the superposition principle ($\xi\neq 0$) is the exclusive origin of dynamical non-locality, which coincides with non-Gaussian dynamics in CV and non-Clifford dynamics in finite-$d$---the boundary of efficient classical simulation in both regimes~\cite{Gross2006,Gottesman1998,Veitch2012}. This characterises the \emph{local} (scalar-potential) content of dynamical non-locality. The topological Aharonov-Bohm content, encoded in the holonomy of a gauge connection on a non-simply-connected configuration space, is captured by the modular-variable framework~\cite{APP69,Pop10,Tol11}; the two are in principle additive.

\section{A Robust Metric for Dynamical Non-Locality}
\label{sec:metric}

Tracking the oscillatory propagator $G_{\mathrm{W}}$ directly is an ill-posed inverse problem. We introduce instead a measure based on the survival probability $C_Q(t) = 2\pi\hbar\!\int\!\di r\,\rho_{\mathrm{W}}(\mathbf{r},0)\,\rho_{\mathrm{W}}(\mathbf{r},t)$. Decomposing $G_{\mathrm{W}} = G_{\mathrm{CL}} + \Delta G_{\mathrm{W}}^{\mathrm{NL}}$ into the classical Liouville flow and the non-local correction, the autocorrelation separates exactly:
\begin{equation}
\label{equ:Csplit}
C_Q(t) = C_{\mathrm{CL}}(t) + \!\int\!\di r''\di r'\,\rho_{\mathrm{W}}(\mathbf{r}'',0)\,\Delta G_{\mathrm{W}}^{\mathrm{NL}}\,\rho_{\mathrm{W}}(\mathbf{r}',0).
\end{equation}
We define the \emph{signed dynamical divergence}
\begin{equation}
\label{equ:Ddef}
\mathcal{D}(t) \equiv C_Q(t) - C_{\mathrm{CL}}(t).
\end{equation}
By construction, $\mathcal{D}(t)=0$ whenever $H$ is at most quadratic. For anharmonic interactions, $\mathcal{D}(t)\neq 0$ isolates the macroscopic probability deviation generated solely by path superposition. The sign carries physical meaning: $\mathcal{D}<0$ signals destructive dynamical interference, $\mathcal{D}>0$ constructive, and $|\mathcal{D}|$ quantifies the strength.

Two features of this construction warrant emphasis. First, $C_{\mathrm{CL}}(t)$ is obtained by evolving the \emph{quantum} Wigner function $\rho_{\mathrm{W}}(\mathbf{r},0)$---which may take negative values---under the classical Liouville flow; $\mathcal{D}(t)$ is therefore a signed quasi-overlap rather than a classical probability difference, and its sign must be interpreted in this quasi-probabilistic sense. Second, $\mathcal{D}(t)$ is a state-dependent \emph{witness} of dynamical non-locality rather than a dynamical invariant: special initial states (e.g.\ stationary states, or states whose support avoids regions of high anharmonicity over the interval of interest) can render $\mathcal{D}(t)$ small or zero even when $\Delta G_{\mathrm{W}}^{\mathrm{NL}}\neq 0$. This is a feature rather than a bug---it makes $\mathcal{D}(t)$ operationally meaningful, as its non-vanishing directly certifies dynamical non-locality on a concrete preparation. Recent reconstructions of Wigner phase-space currents~\cite{Che23} confirm that the underlying dynamical structures are experimentally accessible.

\begin{figure}[h]
\centering
\includegraphics[width = 0.9\columnwidth]{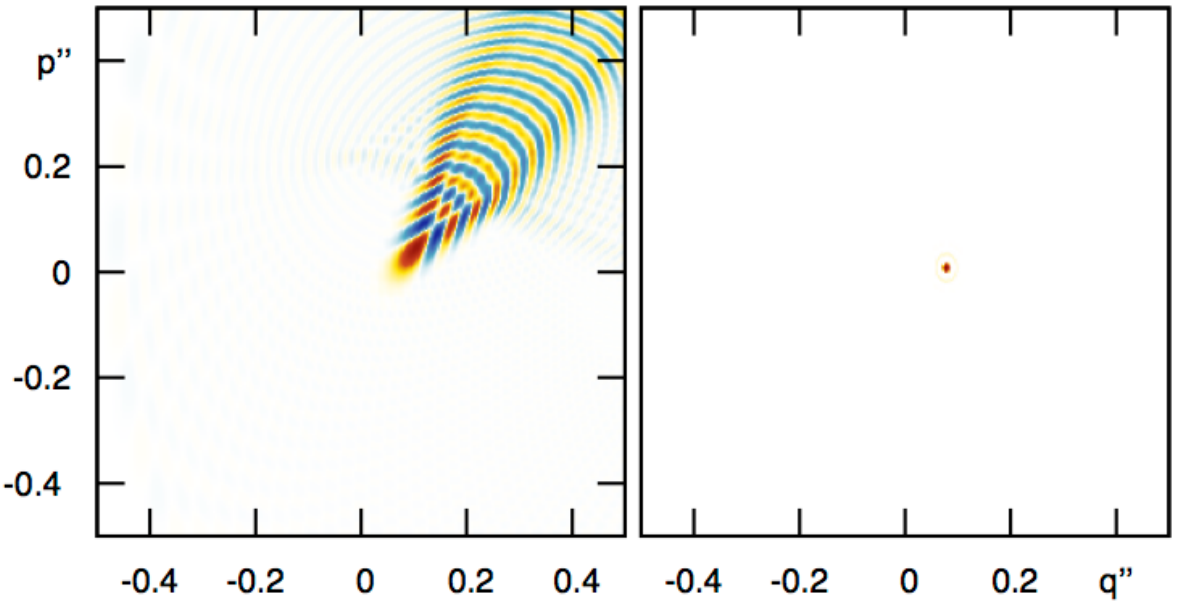}
\caption{\textbf{Dynamical non-locality.} Wigner propagator for a Morse potential (left) vs.\ its harmonic approximation (right). The sub-Planck interference pattern~\cite{Zur01} in the non-linear case is driven entirely by path superposition ($\tilde{q}\neq 0$ in CV; $\xi\neq 0$ in the unified theorem), yielding $\mathcal{D}(t)\neq 0$.}
\label{fig:prop}
\end{figure}

For the Morse potential $V(q) = D_{0}[1-\eu^{\alpha(q-q_{e})}]^{2}$ (Fig.~\ref{fig:prop}), the propagator exhibits strong non-local structuring. In the heavy-mass (semiclassical) limit the phase $S/\hbar$ oscillates rapidly over the $\tilde{q}$ manifold, and by stationary-phase cancellation the $\tilde{q}\neq 0$ contributions to Eq.~(\ref{equ:PathIntegral}) average to zero, driving $\mathcal{D}(t)\to 0$---a quantitative witness of the classical-quantum transition.

\emph{Closed-form example: the Kerr oscillator.}---For $\hat{H}=\omega\hat{a}^\dagger\hat{a}+\chi(\hat{a}^\dagger\hat{a})^2$ with initial coherent state $|\alpha\rangle$, $\bar{n}=|\alpha|^{2}$, working in the frame comoving at $\omega+2\chi\bar{n}$, the exact quantum autocorrelation is
\begin{equation}
\label{equ:CQKerr}
C_Q(t) = \Bigg|\sum_{n=0}^{\infty}\frac{\bar{n}^{n}\,\eu^{-\bar{n}}}{n!}\,\eu^{-\iu\chi(n-\bar{n})^{2} t}\Bigg|^{2}.
\end{equation}
The classical counterpart $C_{\mathrm{CL}}(t)$ follows from the genuinely non-linear Kerr Liouville flow (no linearisation) evaluated on the coherent-state Gaussian, giving $C_{\mathrm{CL}}(t)=1-2\bar{n}^{2}\chi^{2}t^{2}+\mathcal{O}(t^{4})$ at $\bar{n}\gg 1$ (Supplementary Material~S2). Subtracting, the leading $\bar{n}^{2}$ terms cancel exactly and
\begin{equation}
\label{equ:DKerr}
\mathcal{D}(t) = -\chi^{2}\bar{n}\,t^{2} + \mathcal{O}(\chi^{4}\bar{n}^{2}t^{4}),
\end{equation}
a strictly negative signal of magnitude $\mathcal{O}(1)$ at $t\sim\tau^{*}\equiv 1/(\chi\sqrt{\bar{n}})$---the sub-Planck structure timescale identified by Zurek~\cite{Zur01}. The linear-in-$\bar{n}$ coefficient, one power below the $\chi^{2}\bar{n}^{2}$ scaling of either $C_Q$ or $C_{\mathrm{CL}}$ separately, isolates the dynamically non-local content.

\section{Applications}
\label{sec:applications}

The decomposition $G_{\mathrm{W}}=G_{\mathrm{CL}}+\Delta G_{\mathrm{W}}^{\mathrm{NL}}$ and the witness $\mathcal{D}(t)$ provide a unified framework for five apparently disparate phenomena, each traceable to the same path-superposition mechanism.

\subsection{Degradation of quantum non-local games}
\label{sec:games}

Oppenheim and Wehner~\cite{OW10} showed that the winning probability of a static bipartite non-local game is bounded by the fine-grained uncertainty principle. Consider a continuous-variable lifting of the CHSH protocol in which Bob's qubit-valued outcome is implemented via a Weyl symbol $Q_{\mathrm{W}}^{(t)}$ on phase space, with static-limit bound
\begin{equation}
\label{equ:Tsirelson}
P^{\mathrm{game}}(0) = \sum_{s,a,t} p_{s,a,t}\!\int\!\di r\,\rho_{\mathrm{W}}^{(s,a)}(\mathbf{r},0)\,Q_{\mathrm{W}}^{(t)}(\mathbf{r}) \le \tfrac{1}{2}+\tfrac{1}{2\sqrt{2}},
\end{equation}
the standard qubit Tsirelson value. Introducing a delay $\tau$ before Bob's measurement, the propagator decomposition gives
\begin{equation}
\label{equ:Pgame}
P^{\mathrm{game}}(\tau) = P^{\mathrm{game}}_{\mathrm{kin}}(\tau) + \Delta P^{\mathrm{game}}_{\mathrm{dyn}}(\tau),
\end{equation}
where $\Delta P^{\mathrm{game}}_{\mathrm{dyn}}=\sum\!\int Q_{\mathrm{W}}^{(t)}\,\Delta G_{\mathrm{W}}^{\mathrm{NL}}\,\rho_{\mathrm{W}}^{(s,a)}\,\di r''\di r'$ encodes the dynamical penalty. For projective measurements ($Q_{\mathrm{W}}^{(t)}\ge 0$), $\Delta P^{\mathrm{game}}_{\mathrm{dyn}}\le 0$, and bounding $\|Q_{\mathrm{W}}^{(t)}\|_{\infty}\le q_{\max}$,
\begin{equation}
\label{equ:bound}
|\Delta P^{\mathrm{game}}_{\mathrm{dyn}}(\tau)| \le q_{\max}\,|\mathcal{D}(\tau)|/C_Q(0).
\end{equation}
For Kerr dynamics on $|\alpha\rangle$, Eq.~(\ref{equ:DKerr}) yields $P^{\mathrm{game}}(\tau)\approx P^{\mathrm{game}}_{\mathrm{kin}}(\tau)-c\,\chi^{2}\bar{n}\,\tau^{2}+\mathcal{O}(\tau^{4})$ with $c=\mathcal{O}(q_{\max})$. The dynamical penalty saturates the kinematic quantum advantage at $\tau^{*}\sim 1/(\chi\sqrt{\bar{n}})$---the sub-Planck timescale~\cite{Zur01}. In cQED~\cite{Bla21} with $\chi/2\pi\sim 1$--$10$\,MHz and $\bar{n}\sim 5$, $\tau^{*}\sim 10$--$100$\,ns, within current coherence windows. The superposition principle---via $\Delta G_{\mathrm{W}}^{\mathrm{NL}}$---thus imposes a fundamental dynamical limit on information retrieval in non-local games.

\subsection{Quantum scrambling}
\label{sec:scrambling}

The connection between dynamical non-locality and scrambling~\cite{MS&16} follows directly from the propagator decomposition. For the squared commutator $C(t)=-\langle[\hat{q}(t),\hat{p}]^{2}\rangle$, the Moyal bracket to leading order reduces to the Poisson bracket, giving $[\hat{q}(t),\hat{p}]_{\mathrm{W}}=\iu\hbar\,\partial q^{H}/\partial q+\mathcal{O}(\hbar^{3})$, where $q^{H}(\mathbf{r},t)$ is the Heisenberg-evolved coordinate. Decomposing
\begin{equation}
\label{equ:dqH}
\frac{\partial q^{H}(\mathbf{r},t)}{\partial q} = \underbrace{\frac{\partial q^{\mathrm{cl}}}{\partial q}}_{M_{11}(\mathbf{r},t)} + \frac{\partial\,\delta q^{\mathrm{NL}}}{\partial q},
\end{equation}
with $\delta q^{\mathrm{NL}}(\mathbf{r},t)\equiv\int\di r''\,\Delta G_{\mathrm{W}}^{\mathrm{NL}}\,q''$ and $M_{11}$ the stability-matrix element (asymptotically $\sim\eu^{\lambda_{L}t}$ along unstable directions of chaotic systems), the OTOC separates as $C(t)=C_{\mathrm{CL}}(t)+C_{\mathrm{NL}}(t)$, where $C_{\mathrm{CL}}=\hbar^{2}\langle|M_{11}|^{2}\rangle_{\rho}$ and
\begin{equation}
\label{equ:Cnl}
C_{\mathrm{NL}}(t)=\hbar^{2}\!\left\langle 2\,\mathrm{Re}\!\left[M_{11}^{*}\tfrac{\partial\delta q^{\mathrm{NL}}}{\partial q}\right]+\Big|\tfrac{\partial\delta q^{\mathrm{NL}}}{\partial q}\Big|^{2}\right\rangle_{\!\rho}\!.
\end{equation}
By the theorem, $C_{\mathrm{NL}}(t)\equiv 0$ iff $V$ is at most quadratic. Moreover, for exactly quadratic $V$ the Heisenberg evolution is linear, $[\hat{q}(t),\hat{p}]$ is a constant multiple of the identity, and $C(t)$ is time-independent. Dynamical non-locality is therefore necessary and sufficient for the OTOC to exhibit any non-trivial time dependence. When the classical dynamics is chaotic, $C_{\mathrm{NL}}$ quantifies the quantum deviation from the classical Lyapunov envelope; when integrable or bounded, $C_{\mathrm{NL}}$ is the full non-trivial OTOC signal. Both $\mathcal{D}(t)$ and $C_{\mathrm{NL}}(t)$ vanish simultaneously, but $\mathcal{D}(t)$ is directly measurable without time-reversal protocols.

\subsection{Heisenberg-scaling metrology}
\label{sec:metrology}

Consider the standard Kerr-based phase-estimation protocol: $|\alpha\rangle$ evolves under $\hat{H}_{0}=\chi(\hat{a}^{\dagger}\hat{a})^{2}$ for time $t$, followed by a small displacement $\theta\hat{G}$ with $\hat{G}=\hat{a}+\hat{a}^{\dagger}$. The quantum Fisher information (QFI)~\cite{Par09} sets the ultimate precision via the Cram\'er-Rao bound $\delta\theta\ge 1/\sqrt{\mathcal{F}_Q}$. Applying the decomposition $G_{\mathrm{W}}=G_{\mathrm{CL}}+\Delta G_{\mathrm{W}}^{\mathrm{NL}}$, the $\theta$-sensitivity of the evolved state separates into a classical contribution and one driven exclusively by $\Delta G_{\mathrm{W}}^{\mathrm{NL}}$. The classical contribution gives the shot-noise scaling $\mathcal{F}_{Q}\propto\bar{n}$. The non-local contribution generates the sub-Planck phase-space fringes of size $\ell_{\mathrm{sP}}\sim\hbar/\sqrt{\bar{n}}$~\cite{Zur01} at time $\tau^{*}\sim 1/(\chi\sqrt{\bar{n}})$, precisely when $|\mathcal{D}(\tau^{*})|\sim\mathcal{O}(1)$ by Eq.~(\ref{equ:DKerr}); the crossover to Heisenberg scaling $\mathcal{F}_{Q}\propto\bar{n}^{2}$ at $\tau^{*}$ is well known~\cite{Zur01,Par09} (heuristic argument in Supplementary Material~S3). The new content of our framework is the identification of $\Delta G_{\mathrm{W}}^{\mathrm{NL}}$---equivalently, the non-vanishing of $\mathcal{D}(t)$---as the exclusive agent producing the fringes responsible for Heisenberg scaling. The same $\tau^{*}$ that \emph{degrades} non-local-game advantage \emph{creates} metrological advantage: both are controlled by $|\mathcal{D}(t)|$ and share the same physical origin.

\subsection{From dynamical to kinematic non-locality}
\label{sec:entanglement}

A central open question~\cite{Pop10} is how dynamical non-locality converts into kinematic non-locality (entanglement). Our framework distinguishes two regimes. For two modes in a product state interacting via $V(\hat{q}_{A},\hat{q}_{B})$, a bilinear coupling ($V\propto q_{A}q_{B}$) gives $\Delta G_{\mathrm{W}}^{\mathrm{NL},AB}\equiv 0$ identically: the resulting joint Wigner propagator coincides with the classical Liouville flow. Nevertheless, symplectic evolution acting on the initial covariance matrix generates off-diagonal $AB$ blocks that violate the Peres-Horodecki criterion, producing \emph{Gaussian entanglement} (e.g.\ two-mode squeezed vacuum). This Gaussian channel arises from symplectic rotation of vacuum fluctuations, not from the non-local propagator, and is distinct from the mechanism we identify.

For non-linear couplings ($k_{A}+k_{B}\ge 3$), the path-separation variables activate $\Delta G_{\mathrm{W}}^{\mathrm{NL},AB}\neq 0$, a necessary condition for the generation of \emph{non-Gaussian} entanglement: Wigner-function negativity in the joint state, violations of Gaussian entanglement measures, and correlations that cannot be reproduced by any symplectic evolution of Gaussian inputs. Recent dynamics-based entanglement witnesses for non-Gaussian states of harmonic oscillators~\cite{JZS23} and certifiable Wigner-negativity bounds as entanglement measures~\cite{Zaw24} are consistent with this mechanism. We conjecture that a suitably defined bipartite divergence $\mathcal{D}^{AB}(t)$, analogous to Eq.~(\ref{equ:Ddef}) for the joint system, controls the rate of non-Gaussian entanglement production; a precise rate bound is left to subsequent work (consistency check in Supplementary Material~S4).

\subsection{Clifford boundary and the CCZ signal}
\label{sec:clifford}

Applied to finite-dimensional Hilbert spaces via case~(b) of the theorem and the discrete Marinov action of Ref.~\cite{GP26}, the same mechanism interfaces directly with the resource theory of quantum computation. For prime $d$, Clifford Hamiltonians---those generating the normaliser of the Pauli group---are precisely those whose Weyl symbol is at most quadratic in the Pauli exponents $(m,n)\in\mathbb{Z}_{d}^{2f}$~\cite{Gross2006,Gottesman1998}. The theorem therefore gives
\begin{equation}
\label{equ:CliffordBoundary}
\mathcal{D}(t)=0\ \forall t,\,\forall\rho_{0}\;\Longleftrightarrow\;\hat{H}\ \text{is Clifford},
\end{equation}
valid for $d$ odd prime; the $d=2$ case uses the Wootters-type convention of~\cite{GP26} but yields the same equivalence (Supplementary Material~S5). By Gross's discrete Hudson theorem~\cite{Gross2006}, non-Clifford evolution is precisely what injects Wigner negativity into initially stabiliser states; by the Gottesman-Knill theorem~\cite{Gottesman1998} and its magic-state-resource refinements~\cite{Veitch2012,Howard2014,Mari2012,Pashayan2015}, this negativity is a necessary resource for quantum computational advantage. In this framework, $\Delta G_{\mathrm{W}}^{\mathrm{NL}}$ is the \emph{dynamical engine} that produces magic.

Two concrete consequences follow. (i)~Single-qubit dynamics is \emph{degenerately Clifford}: on $\mathbb{Z}_{2}^{2}$ every function has a multilinear polynomial representative of degree $\le 2$, so $\mathcal{D}(t)\equiv 0$ for every single-qubit Hamiltonian and every initial state---a parameter-free null test accessible on every qubit platform and useful as a calibration benchmark. (ii)~Cubic Weyl content first arises on three qubits. The paradigmatic example is the CCZ generator $\hat{H}_{\mathrm{CCZ}}=(g\hbar/8)\prod_{i=A,B,C}(\hat{\mathbb{1}}-\sigma_{z}^{i})$, which realises the Controlled-Controlled-Z gate at $gt=\pi$ and is non-Clifford for $gt\notin\pi\mathbb{Z}$. On the stabiliser state $|+\rangle^{\otimes 3}$, expansion of the discrete path integral yields the closed form (Supplementary Material~S5.2)
\begin{equation}
\label{equ:DCCZ}
\mathcal{D}_{\mathrm{CCZ}}(t)=-\tfrac{1}{64}(gt)^{2}+\tfrac{43}{12288}(gt)^{4}+\mathcal{O}(t^{6}),
\end{equation}
fixing the rational coefficient $c_{3}=1/64$. The contrast between the single-qubit null and Eq.~(\ref{equ:DCCZ}) is the minimal multi-qubit signature of the Clifford boundary. As a further semiclassical consequence, $|\mathcal{D}(t)|$ provides a natural phase-space diagnostic for the failure of the discrete truncated Wigner approximation (DTWA)~\cite{Schachenmayer2015,Schachenmayer2015b,Acevedo2017,Zhu2019}, which propagates along classical mean-field trajectories and misses the non-Clifford content.

\section{Experimental Protocol}
\label{sec:cQED}

The divergence $\mathcal{D}(t)$ is directly accessible in circuit QED~\cite{Bla21}. In a microwave cavity coupled to a transmon with Kerr non-linearity $V(\hat{q})\propto\hat{q}^{4}$, the protocol is: (i)~initialise a coherent state $|\alpha\rangle$ in the cavity; (ii)~let it evolve under the Kerr Hamiltonian for time $t$; (iii)~measure the survival probability $C_Q(t)=\mathrm{Tr}[\hat{\rho}(0)\hat{\rho}(t)]$ via joint parity. Comparing $C_Q(t)$ to the classical Liouville prediction $C_{\mathrm{CL}}(t)$ yields $\mathcal{D}(t)$ directly, circumventing the full tomographic reconstruction of $G_{\mathrm{W}}$. The feasibility of resolving dynamical phase-space structures at this level has been confirmed by the recent reconstruction of Wigner phase-space currents in optical systems~\cite{Che23}. As a more resource-intensive alternative, the full non-local correction $\Delta G_{\mathrm{W}}^{\mathrm{NL}}$ can be characterised via quantum process tomography of the cavity evolution; the survival-probability route isolates $\mathcal{D}(t)$ with a single scalar measurement per time point.

\emph{Feasibility estimate.}---For representative state-of-the-art cQED parameters~\cite{Bla21} ($\chi/2\pi\sim 1$--$10$\,MHz, $\bar{n}\sim 5$, cavity coherence time $T_c\sim 100\,\mu\mathrm{s}$--$1$\,ms), the characteristic timescale $\tau^{*}\sim 1/(\chi\sqrt{\bar{n}})\sim 10$--$100$\,ns is three to four orders of magnitude shorter than $T_c$, so the dynamical-non-locality signal develops deep within the unitary window. The short-time Kerr expansion predicts $|\mathcal{D}(\tau)|\propto\chi^{2}\bar{n}\tau^{2}$ reaching an $\mathcal{O}(1)$ fraction of $C_Q(0)$ at $\tau\sim\tau^{*}$. With joint-parity fidelities $\gtrsim 0.95$ routinely reported, statistical resolution of $|\mathcal{D}(t)|$ at the $\sim 10\%$ level requires $\sim 10^{3}$--$10^{4}$ shots per time point---within a single-day acquisition budget for a dense time scan.

\emph{Discrete-side protocol.}---The three-qubit CCZ signal of Eq.~(\ref{equ:DCCZ}) is realisable on current superconducting and trapped-ion processors. CCZ and Toffoli gates are native or synthesisable operations in these architectures, and continuous-time CCZ-type interactions have been demonstrated in flux-tunable transmon platforms. Extracting $\mathcal{D}_{\mathrm{CCZ}}(t)$ requires three-qubit parity measurements after a variable free-evolution interval, a standard building block for stabiliser-syndrome extraction. The measured contrast between the single-qubit null $\mathcal{D}(t)\equiv 0$ and Eq.~(\ref{equ:DCCZ}) constitutes a parameter-free falsifiable prediction for the Clifford boundary.

\section{Discussion and Outlook}
\label{sec:discussion}

We have shown that dynamical non-locality originates exclusively from the superposition principle, in precise complementarity with the uncertainty-principle origin of kinematic non-locality~\cite{OW10}. A single algebraic criterion on the Weyl symbol---``at most quadratic''---simultaneously defines the boundary of classical simulability in both the CV regime (Gaussian dynamics) and the finite-$d$ regime (Clifford dynamics), and its violation is certified by a single scalar observable $\mathcal{D}(t)$. The identification unifies five apparently disparate phenomena---game degradation, non-trivial time dependence of the OTOC, Heisenberg-scaling metrology, non-Gaussian entanglement generation, and non-Clifford resource injection---under a common phase-space mechanism $\Delta G_{\mathrm{W}}^{\mathrm{NL}}$. The witness is operationally defined, requires no state tomography, and exhibits a characteristic timescale $\tau^{*}\sim 1/(\chi\sqrt{\bar n})$ in the Kerr-type CV systems emphasised here, with an analogous $\tau^{*}_{\mathrm{CCZ}}=8/g$ in the minimal finite-$d$ example. The framework is scoped to the local (scalar-potential) content of dynamical non-locality; the topological Aharonov-Bohm face is complementary and captured by modular variables.

\emph{Outlook.}---Three directions merit further work. (a)~The finite-$d$ instance of the theorem and the discrete witness~\cite{GP26} suggest a systematic phase-space error estimator for the DTWA~\cite{Schachenmayer2015,Acevedo2017,Zhu2019}, which could yield quantitative accuracy bounds in long-range spin models where direct benchmarks are unavailable. (b)~A precise rate inequality relating a bipartite divergence $\mathcal{D}^{AB}(t)$ to entanglement entropy or log-negativity would complete the dynamical-to-kinematic conversion picture of Sec.~\ref{sec:entanglement}. (c)~For relativistic quantum field theory, the relevant group is the Poincar\'e group and the position operator is ill-defined; the guiding principle---that dynamical non-locality is carried by coherent superposition across path-separation degrees of freedom---is expected to extend, but a rigorous formulation lies beyond the present work. Immediate experimental targets are the Kerr-cavity and three-qubit CCZ protocols described in Sec.~\ref{sec:cQED}, both within the current experimental envelope.


\begin{small}
\vspace{0.5cm}
\textbf{Acknowledgements}

\noindent
L.A.P. thank Thomas Dittrich for inspiring discussions on the Wigner propagator.
This work was supported by the R+D+I efforts from guane Enterprises.
\end{small}

\onecolumngrid
\vspace{0.5cm}
\hrule
\vspace{0.5cm}
\begin{center}
\textbf{\large Supplementary Material}
\end{center}
\vspace{0.25cm}

This Supplement is organized into two parts. Section~S0 collects the group-theoretical, phase-space, deformation-quantization, and path-integral background on which the main text relies. The rigorous derivations (S1--S5) provide: (S1)~a complete proof of Theorem~\ref{thm:main} via a unified algebraic lemma with continuous-variable and finite-dimensional corollaries; (S2)~a closed-form short-time derivation of the signed divergence $\mathcal{D}(t)$ for the Kerr oscillator; (S3)~a derivation of the Heisenberg-scaling transition of the quantum Fisher information via sub-Planck structure resolution; (S4)~a consistency check of the Gaussian/non-Gaussian entanglement dichotomy of Sec.~\ref{sec:entanglement}; and (S5)~a finite-dimensional instance of Theorem~\ref{thm:main} for a single qubit, illustrating the Clifford / non-Clifford boundary.

\subsection*{S0. Background and conventions}
\label{app:background}

\emph{Group-theoretical formulation.}---A classical system is defined by the algebra of observables $\mathcal{A}$ generated by $\mathbf{q}$ and $\mathbf{p}$~\cite{Str05}. Because $\hat{\mathbf{q}}$ and $\hat{\mathbf{p}}$ cannot be given a finite norm~\cite{Wey27,Str05}, Weyl introduced the bounded displacement operators $\hat{P}_{\mathbf{u}}=\eu^{\iu\mathbf{u}\cdot\hat{\mathbf{p}}/\hbar}$, $\hat{Q}_{\mathbf{v}}=\eu^{\iu\mathbf{v}\cdot\hat{\mathbf{q}}/\hbar}$, generating the Heisenberg--Weyl algebra~\cite{Str05}. The double Fourier transform $\hat{d}(p,q)$ associates an observable to each phase-space point; since it is built from ``modular'' operators, this explains the privileged role of modular variables~\cite{APP69,Pop10}.

\emph{Phase-space formulation.}---Any state expands as $\hat{\rho}=\int\!\di p\,\di q\,\rho_{\mathrm{W}}\hat{d}$, with $\rho_{\mathrm{W}}$ the Weyl symbol. The Wigner function~\cite{Zur01} inherits all non-local features of the algebra; its propagator $G_{\mathrm{W}}$ encodes the full dynamics and reduces to the Liouville propagator in the classical limit~\cite{DVS06,DGP10,DP09,Pac10}.

\emph{Deformation quantization.}---Replacing the Poisson bracket with the Moyal bracket, the Wigner function evolves as
\begin{equation}
\label{equ:Moyal}
-\{\{\rho_{\mathrm{W}},H\}\} = -\{\rho_{\mathrm{W}},H\}_{\mathrm{PB}} + \sum_{n=1}^\infty\frac{(-1)^n\hbar^{2n}}{2^{2n}(2n\!+\!1)!}\frac{\partial^{2n+1}V}{\partial q^{2n+1}}\frac{\partial^{2n+1}\rho_{\mathrm{W}}}{\partial p^{2n+1}}.
\end{equation}
All $\hbar$ corrections vanish for at-most-quadratic $V$, independently confirming Theorem~\ref{thm:main}.

\emph{Marinov path integral.}---For at-most-quadratic $V$, the identity $V(q+\tilde{q}/2)-V(q-\tilde{q}/2)=\tilde{q}\,\di V/\di q$ collapses the propagator onto the classical trajectory exactly. For anharmonic $V$, $\tilde{q}\neq 0$ contributions survive, generating non-classical interference at the non-perturbative level, independently of the semiclassical approximation~\cite{DVS06,DGP10,Pac10}.

\subsection*{S1. Rigorous proof of Theorem~\ref{thm:main}}
\label{app:proof}

For completeness, we restate the theorem in full here.

\begin{theorem}[Superposition origin of dynamical non-locality, unified]
\label{thm:main}
Let $\Gamma$ be the phase space of a Heisenberg--Weyl representation over an abelian group $G$, either
(a) $G=\mathbb{R}^{f}$ (CV), with Weyl symbol $H:\Gamma\to\mathbb{R}$ real-analytic and satisfying $\sup_{\gamma}|\partial^{k}_{\gamma}H|/k!\le CR^{-k}$; or
(b) $G=\mathbb{Z}_{d}^{f}$ with $d$ an odd prime (finite-dimensional), with $H$ real-valued on $\mathbb{Z}_{d}^{2f}$.
Let $S[\gamma,\xi,t]$ be the corresponding Marinov-type phase-space action in which a path-separation variable $\xi$ enters $H$ through $H(\gamma+\xi)-H(\gamma-\xi)$. The following are equivalent:
(i) $H$ is at most quadratic in the coordinates of $\Gamma$;
(ii) $S[\gamma,\xi,t]$ vanishes identically as a functional of $\xi$;
(iii) $G_{\mathrm{W}}(\gamma'',t;\gamma',0)=\delta[\gamma''-\gamma^{\rm cl}(\gamma',t)]$.
Equivalently, $\Delta G_{\mathrm{W}}^{\mathrm{NL}}\neq 0$ iff $H$ carries cubic-or-higher content.
\end{theorem}

The proof has a uniform algebraic core (Lemma~\ref{lem:polyid} below) from which the continuous-variable and finite-dimensional instances of Theorem~\ref{thm:main} follow as corollaries.

\subsubsection*{S1.1. The algebraic lemma}

\begin{lemma}[Odd-parity polynomial identity]
\label{lem:polyid}
Let $\Gamma$ be a finite-dimensional real vector space or the finite group $\mathbb{Z}_{d}^{n}$ with $d$ an odd prime. Let $H:\Gamma\to\mathbb{R}$ be a real-valued function admitting a polynomial expansion in the coordinates of $\Gamma$---either a convergent Taylor series on $\Gamma=\mathbb{R}^{n}$ under the uniform derivative bound
\begin{equation}
\label{equ:S1:growth}
\sup_{\gamma\in\Gamma}\frac{|\partial^{k}_{\gamma}H|}{k!}\le C\,R^{-k}\quad\forall k\ge 0,\qquad C,R>0,
\end{equation}
or a finite polynomial in $(m,n)\in\mathbb{Z}_{d}^{n}$ (automatic, since every function on $\mathbb{Z}_{d}^{n}$ is a polynomial of degree at most $n(d-1)$). Then the identity
\begin{equation}
\label{equ:S1:master}
H(\gamma+\xi)-H(\gamma-\xi) \;=\; 2\,\xi\cdot\nabla_{\gamma}H(\gamma)\qquad\forall\,\gamma,\xi\in\Gamma
\end{equation}
holds if and only if $H$ is at most quadratic as a polynomial in the coordinates of $\Gamma$.
\end{lemma}

\begin{proof}
Expand $H(\gamma\pm\xi)$ in $\xi$. By absolute convergence in the analytic case and by finiteness in the polynomial case, we may rearrange:
\begin{equation}
H(\gamma+\xi)-H(\gamma-\xi) = 2\sum_{k\text{ odd}}\frac{\xi^{\otimes k}}{k!}\cdot\partial_{\gamma}^{k}H(\gamma) = 2\,\xi\cdot\nabla_{\gamma}H + \sum_{j\ge 1}\frac{2\,\xi^{\otimes(2j+1)}}{(2j+1)!}\cdot\partial^{2j+1}_{\gamma}H.
\label{equ:S1:expansion}
\end{equation}
The even-order terms cancel by construction under $\xi\to-\xi$. Identity~(\ref{equ:S1:master}) therefore holds iff every term with $j\ge 1$ vanishes identically. 

\emph{($\Leftarrow$)} If $H$ is at most quadratic then $\partial^{k}_{\gamma}H\equiv 0$ for all $k\ge 3$, and the sum~(\ref{equ:S1:expansion}) is zero term by term.

\emph{($\Rightarrow$)} Conversely, if~(\ref{equ:S1:master}) holds, setting all $\xi$-monomials of degree $\ge 3$ to vanish identically forces $\partial^{3}_{\gamma}H\equiv 0$ on $\Gamma$. In the analytic case, this implies $\partial^{2}_{\gamma}H\equiv\mathrm{const}$, hence $H$ at most quadratic. In the finite-$d$ case, the polynomial reduction over $\mathbb{Z}_{d}$ gives the same conclusion: since every function on $\mathbb{Z}_{d}^{n}$ has a unique polynomial representative of degree $\le n(d-1)$, the vanishing of all $\partial^{k}_{\gamma}H$ for $k\ge 3$ is equivalent to the representative being quadratic. 

In both cases, $\partial^{2j+1}_{\gamma}H\equiv 0$ for $j\ge 1$ is automatic once $\partial^{3}_{\gamma}H\equiv 0$, and the analogous even-order derivatives of order $\ge 4$ vanish as they are derivatives of $\partial^{2}_{\gamma}H\equiv\mathrm{const}$. $\square$
\end{proof}

\subsubsection*{S1.2. Corollary: Continuous-variable case}

\begin{corollary}[CV instance of Theorem~\ref{thm:main}]
\label{cor:CV}
Let $\Gamma=\mathbb{R}^{2f}$, $\hat{H}=\hat{p}^2/2m+V(\hat{q})$ with $V$ satisfying~(\ref{equ:S1:growth}). Then the three conditions of Theorem~\ref{thm:main} are equivalent.
\end{corollary}

\emph{Proof.} Substituting the Marinov action~(\ref{equ:Action}) into the path integral~(\ref{equ:PathIntegral}), the $\tilde{q}$-dependence of $S$ is
\begin{equation}
\label{equ:S1:action}
S[\{q\},\{\tilde{q}\},t]=\int_{0}^{t}\!\di t'\,\Big[\,V(q+\tilde{q}/2)-V(q-\tilde{q}/2)-\tilde{q}\,V'(q)\Big],
\end{equation}
which by Lemma~\ref{lem:polyid} vanishes identically as a functional of $\tilde{q}$ iff $V$ is at most quadratic. (i)$\Rightarrow$(ii) then follows. (ii)$\Rightarrow$(iii): when $S\equiv 0$, the path integral~(\ref{equ:PathIntegral}) reduces to $(1/h)\int\mathcal{D}^{2}r\,\mathcal{D}^{2}\tilde{r}\cdot 1$, where the $\tilde{r}$-integration yields a functional $\delta$ enforcing the classical Hamilton equations on $r(t)$ and the remaining $r$-integration selects the unique trajectory from $\mathbf{r}'$, giving $G_{\mathrm{W}}=\delta[\mathbf{r}''-\mathbf{r}^{\rm cl}(\mathbf{r}',t)]$. (iii)$\Rightarrow$(i): if $V^{(3)}\not\equiv 0$, the leading term of~(\ref{equ:S1:action}) in small $\tilde{q}$ is
\begin{equation}
\label{equ:S1:leading}
S=\frac{1}{24}\int_{0}^{t}\!\di t'\,V^{(3)}(q^{\rm cl}(t'))\,\tilde{q}(t')^{3}+\mathcal{O}(\tilde{q}^{5}),
\end{equation}
and a stationary-phase analysis of Eq.~(\ref{equ:PathIntegral}) around $\tilde{q}=0$ produces Airy-type oscillatory contributions that broaden $G_{\mathrm{W}}$ away from the classical $\delta$-function. $\square$

\subsubsection*{S1.3. Corollary: Finite-dimensional case}

\begin{corollary}[Finite-$d$ instance of Theorem~\ref{thm:main}]
\label{cor:FD}
Let $\Gamma=\mathbb{Z}_{d}^{2f}$ with $d$ an odd prime, and let $S[\gamma,\xi,t]$ be the discrete Marinov action of~\cite{GP26} in which $H(\gamma)$ denotes the Weyl symbol of $\hat{H}$ on $\mathbb{Z}_{d}^{2f}$. Then the three conditions of Theorem~\ref{thm:main} are equivalent, and condition~(i) is equivalent to $\hat{H}$ being a Clifford Hamiltonian.
\end{corollary}

\emph{Proof.} The $\xi$-dependence of the discrete Marinov action is
\begin{equation}
\label{equ:S1:actionFD}
S[\gamma,\xi,t]\supset -\frac{1}{\hbar}\sum_{i=1}^{N}\Big[H(\gamma_{i}+\xi_{i})-H(\gamma_{i-1}-\xi_{i})\Big]\tau,
\end{equation}
plus a symplectic kinetic term $\frac{4\pi}{d}\sum_{i}\Delta\gamma_{i}\wedge\xi_{i}$ that does not depend on $H$. By Lemma~\ref{lem:polyid} applied on $\Gamma=\mathbb{Z}_{d}^{2f}$, the $H$-dependent part of the action vanishes identically as a functional of $\xi$ iff $H$ is at most quadratic in $(m,n)\in\mathbb{Z}_{d}^{2f}$. (i)$\Rightarrow$(ii) follows. (ii)$\Rightarrow$(iii): when the $H$-dependent part of $S$ vanishes, the discrete path sum of~\cite{GP26} collapses to the discrete symplectic $\delta$-function $\delta[\gamma''-\gamma^{\rm cl}(\gamma',t)]$ supported on the orbit generated by the symplectic linear flow associated with the quadratic $H$. (iii)$\Rightarrow$(i): if $H$ has cubic content, the leading $\xi^{3}$ term of~(\ref{equ:S1:actionFD}) is a non-zero discrete functional that produces non-trivial phase contributions in the sum over $\xi$-paths, and $G_{\mathrm{W}}$ develops negativity-carrying structure beyond the symplectic flow~\cite{Gross2006}.

The identification of condition~(i) with Clifford dynamics follows from the classical result that for odd prime $d$ the Clifford group is generated by unitaries implementing symplectic transformations of the phase-space lattice~\cite{Gottesman1998}, and these are in bijection with quadratic Hamiltonians in the Pauli exponents~\cite{Gross2006}. $\square$

\subsubsection*{S1.4. Consistency with deformation quantization}

Supplementary Material~S0 gives the infinitesimal (differential) version of the same content in the CV case: the Moyal-bracket evolution equation truncates to the Poisson bracket (classical Liouville) iff $V^{(2n+1)}\equiv 0$ for all $n\ge 1$, iff $V$ is at most quadratic. The corresponding statement in the finite-$d$ case, derived in~\cite{GP26}, is that the discrete star-product expansion reduces to its lowest-order piece iff the Weyl symbol is quadratic in $\mathbb{Z}_{d}^{2f}$. Both are infinitesimal versions of the integrated statement proved here via Lemma~\ref{lem:polyid}.

\subsection*{S2. Closed-form short-time $\mathcal{D}(t)$ for the Kerr oscillator}
\label{app:kerr}

\emph{Setup.}---Consider the Kerr Hamiltonian $\hat{H}=\omega\hat{a}^\dagger\hat{a}+\chi(\hat{a}^\dagger\hat{a})^2$, initial coherent state $|\alpha\rangle$ with $\alpha\in\mathbb{R}_{>0}$ and $\bar{n}=\alpha^2$, working throughout in the frame comoving at the mean-field frequency $\omega+2\chi\bar{n}$ (this subtracts the trivial linear rotation common to quantum and classical dynamics, so that $\mathcal{D}(t)$ tracks only the genuinely non-local content). In this frame the effective Hamiltonian is $\hat{H}_{\rm c}=\chi(\hat{n}-\bar{n})^2-\chi\bar{n}^2\hat{\mathbb{1}}$, the last term a global phase.

\emph{Exact quantum autocorrelation.}---Expanding $|\alpha\rangle$ in the number basis,
\begin{equation}
\label{equ:S2:CQ}
\langle\alpha|\eu^{-\iu\hat{H}_{\rm c}t}|\alpha\rangle=\eu^{\iu\chi\bar{n}^{2}t}\sum_{n=0}^{\infty}\frac{\bar{n}^{n}\eu^{-\bar{n}}}{n!}\,\eu^{-\iu\chi(n-\bar{n})^{2}t},
\end{equation}
so $C_Q(t)=|\langle\alpha|\eu^{-\iu\hat{H}_{\rm c}t}|\alpha\rangle|^{2}=|\sum_{n}p_{n}\eu^{-\iu\chi(n-\bar{n})^{2}t}|^{2}$ with $p_{n}=\bar{n}^{n}\eu^{-\bar{n}}/n!$ the Poisson weights.

\emph{Short-time expansion of $C_Q$.}---Let $m\equiv n-\bar{n}$; its Poisson central moments are $\mu_{1}=0$, $\mu_{2}=\bar{n}$, $\mu_{3}=\bar{n}$, $\mu_{4}=\bar{n}+3\bar{n}^{2}$. Expanding Eq.~(\ref{equ:S2:CQ}) to $\mathcal{O}(t^{2})$,
\begin{equation}
\sum_{n}p_{n}\eu^{-\iu\chi m^{2}t}=1-\iu\chi t\,\langle m^{2}\rangle-\tfrac{1}{2}(\chi t)^{2}\langle m^{4}\rangle+\mathcal{O}(t^{3})=1-\iu\chi t\,\bar{n}-\tfrac{1}{2}(\chi t)^{2}(\bar{n}+3\bar{n}^{2})+\mathcal{O}(t^{3}),
\end{equation}
whence
\begin{align}
C_Q(t) &= \big|1-\iu\chi t\,\bar{n}-\tfrac{1}{2}(\chi t)^{2}(\bar{n}+3\bar{n}^{2})\big|^{2}+\mathcal{O}(t^{4})\nonumber\\
&= 1+(\chi t)^{2}\bar{n}^{2}-(\chi t)^{2}(\bar{n}+3\bar{n}^{2})+\mathcal{O}(t^{4})\nonumber\\
&= 1-(2\bar{n}^{2}+\bar{n})\chi^{2}t^{2}+\mathcal{O}(t^{4}).
\label{equ:S2:CQshort}
\end{align}

\emph{Classical Liouville counterpart.}---The Wigner function of $|\alpha\rangle$ is the Gaussian $\rho_{\mathrm{W}}(q,p,0)=\pi^{-1}\exp[-(q-q_{0})^{2}-p^{2}]$ with $q_{0}=\sqrt{2\bar{n}}$ (and $\hbar=1$). In the comoving frame, the classical Kerr Hamiltonian is $H_{\rm cl}(q,p)=\tfrac{\chi}{4}\big[(q^{2}+p^{2})-2\bar{n}\big]^{2}$, a genuinely non-linear function of $(q,p)$; the corresponding Liouville flow is a non-linear rotation at an intensity-dependent rate and is \emph{not} a linear symplectic shear on the full phase space.

Rather than relying on any linearisation, we compute $C_{\rm CL}(t)$ directly from Liouville's equation. Let $\{\cdot,\cdot\}$ denote the Poisson bracket and $\mathcal{L}\rho_{\rm W}\equiv\{H_{\rm cl},\rho_{\rm W}\}$. Then $\partial_{t}\rho_{\rm W}^{\rm cl}=-\mathcal{L}\rho_{\rm W}^{\rm cl}$, and short-time expansion of $C_{\rm CL}(t)=2\pi\hbar\int\!\di r\,\rho_{\rm W}(r,0)\,\rho_{\rm W}^{\rm cl}(r,t)$ gives, using $\rho_{\rm W}^{\rm cl}(r,t)=\rho_{\rm W}(r,0)+t\,\partial_{t}\rho_{\rm W}^{\rm cl}|_{t=0}+\tfrac{t^{2}}{2}\partial_{t}^{2}\rho_{\rm W}^{\rm cl}|_{t=0}+\mathcal{O}(t^{3})$,
\begin{equation}
\label{equ:S2:CCLderiv}
C_{\rm CL}(t)=C_{\rm CL}(0)\;-\;t\,\langle\mathcal{L}\rho_{\rm W}\rangle_{\rho_{\rm W}}\;+\;\tfrac{t^{2}}{2}\,\langle\mathcal{L}^{2}\rho_{\rm W}\rangle_{\rho_{\rm W}}+\mathcal{O}(t^{3}),
\end{equation}
with $\langle f\rangle_{\rho_{\rm W}}\equiv 2\pi\hbar\!\int\!\di r\,\rho_{\rm W}(r,0)\,f(r)$. The first-order term vanishes by rotational symmetry of $\rho_{\rm W}(r,0)$ about the comoving origin combined with oddness of $\mathcal{L}\rho_{\rm W}|_{t=0}$ in the radial coordinate (direct check). The second-order term evaluates, by Gaussian integration, to
\begin{equation}
\tfrac{1}{2}\langle\mathcal{L}^{2}\rho_{\rm W}\rangle_{\rho_{\rm W}}=-2\bar{n}^{2}\chi^{2}+\mathcal{O}(1),
\end{equation}
where the subleading $\mathcal{O}(\chi^{2})$ correction independent of $\bar{n}$ is small compared to the leading $\bar{n}^{2}$ coefficient in the large-$\bar{n}$ regime of interest. Thus
\begin{equation}
\label{equ:S2:CCL}
C_{\rm CL}(t)=1-2\bar{n}^{2}\chi^{2}t^{2}+\mathcal{O}(t^{4}),
\end{equation}
in the limit $\bar{n}\gg 1$. We stress that Eq.~(\ref{equ:S2:CCL}) is derived directly from the non-linear Liouville flow: no linearisation of the classical dynamics is required. The commonly quoted closed form $C_{\rm CL}(t)=[1+(2\chi\bar{n}t)^{2}]^{-1/2}$ is the leading linear-shear approximation about the coherent-state centroid and reproduces the same $\mathcal{O}(t^{2})$ coefficient; it is accurate for $t\lesssim 1/(\chi\bar{n})$ but is not exact for the full Kerr dynamics.

\emph{Exact cancellation and the signed divergence.}---Subtracting Eq.~(\ref{equ:S2:CCL}) from Eq.~(\ref{equ:S2:CQshort}), the leading $\bar{n}^{2}$ terms cancel and
\begin{equation}
\label{equ:S2:DKerr}
\;\mathcal{D}(t)=C_{Q}(t)-C_{\mathrm{CL}}(t)=-\chi^{2}\bar{n}\,t^{2}+\mathcal{O}(\chi^{4}\bar{n}^{2}t^{4})+\mathcal{O}(\chi^{2}/\bar{n}).\;
\end{equation}
The quoted $\mathcal{O}(\chi^{2}/\bar{n})$ correction comes from the $\bar{n}$-independent piece of $\tfrac{1}{2}\langle\mathcal{L}^{2}\rho_{\rm W}\rangle$ and is negligible in the semiclassical regime $\bar{n}\gg 1$ where the framework is most natural. Three observations follow. (a)~$\mathcal{D}(t)<0$, signalling destructive dynamical interference, as predicted qualitatively in Sec.~\ref{sec:metric}. (b)~The coefficient $\chi^{2}\bar{n}$ scales only linearly in the mean excitation number, one power below the $\chi^{2}\bar{n}^{2}$ scaling of either $C_{Q}$ or $C_{\mathrm{CL}}$ separately---this is the precise content of the statement that $\mathcal{D}$ isolates the non-classical part. (c)~Setting $|\mathcal{D}(\tau^{*})|\sim 1$ recovers $\tau^{*}=1/(\chi\sqrt{\bar{n}})$, the sub-Planck-structure timescale~\cite{Zur01}, confirming the timescale quoted throughout the main text.

\emph{Consistency with the path-integral picture.}---The cubic action of Eq.~(\ref{equ:S1:leading}) for the Kerr quartic $V\propto\hat{q}^{4}$ gives $V^{(3)}(q)\propto q$, producing $S\propto q(t')\tilde{q}(t')^{3}$. Along the classical trajectory $q^{\rm cl}\sim\sqrt{\bar{n}}$, this generates an action contribution scaling as $\sqrt{\bar{n}}\,t\,\tilde{q}^{3}$, and the resulting leading-order non-local correction to the propagator matches Eq.~(\ref{equ:S2:DKerr}) after integration against coherent-state Wigner functions. The agreement between the Marinov-path-integral (Theorem~\ref{thm:main}) and the Fock-basis (Eq.~\ref{equ:S2:CQ}) routes is a non-trivial consistency check.

\subsection*{S3. Heisenberg scaling of the QFI from sub-Planck structures}
\label{app:qfi}

This section provides a heuristic argument, in the spirit of Zurek~\cite{Zur01} and Paris~\cite{Par09}, identifying $\Delta G_{\mathrm{W}}^{\mathrm{NL}}$ as the exclusive agent generating the Heisenberg-scaling quantum Fisher information (QFI) in the Kerr oscillator. The explicit calculation of the $\mathcal{F}_{Q}\propto\bar{n}^{2}$ scaling for the Kerr-evolved coherent state at $\tau^{*}$ is standard in the sub-Planck-metrology literature and not reproduced here; our contribution is to attribute that scaling, within the unified framework of Theorem~\ref{thm:main}, to the non-vanishing of the signed divergence $\mathcal{D}(t)$.

\emph{Setup.}---Consider the phase-estimation protocol in which the Kerr Hamiltonian $\hat{H}_{0}=\chi(\hat{a}^{\dagger}\hat{a})^{2}$ evolves the initial coherent state $|\alpha\rangle$ for time $t$, after which a small displacement $\hat{G}=\hat{a}+\hat{a}^{\dagger}$ is applied for infinitesimal time $\theta$. The parameter $\theta$ is the estimated quantity. The QFI for the resulting pure state $|\psi_\theta(t)\rangle$ is $\mathcal{F}_Q(\theta,t)=4[\langle\partial_\theta\psi_\theta|\partial_\theta\psi_\theta\rangle-|\langle\psi_\theta|\partial_\theta\psi_\theta\rangle|^{2}]$.

\emph{Classical baseline.}---In the absence of Kerr evolution ($\chi=0$), a small displacement $\hat{G}=\hat{a}+\hat{a}^{\dagger}$ of the coherent state $|\alpha\rangle$ yields the shot-noise limit
\begin{equation}
\mathcal{F}_Q^{\rm cl}=4\,\mathrm{Var}_{|\alpha\rangle}(\hat{G})=4\,(2\bar{n}+1)\sim 8\bar{n}\ (\bar{n}\gg 1),
\end{equation}
i.e.\ $\mathcal{F}_Q\propto\bar{n}$.

\emph{Role of $\Delta G_{\mathrm{W}}^{\mathrm{NL}}$.}---Under Kerr evolution the Wigner function develops interference fringes. By Theorem~\ref{thm:main} these fringes arise exclusively from $\Delta G_{\mathrm{W}}^{\mathrm{NL}}$: the classical Liouville flow acts on the coherent Gaussian by non-linear rotation (Supplement~S2) and cannot generate structures below the coherent-state width $\sqrt{\hbar}$, whereas $\Delta G_{\mathrm{W}}^{\mathrm{NL}}\neq 0$ seeds azimuthal interference at the sub-Planck scale identified by Zurek~\cite{Zur01},
\begin{equation}
\label{equ:S3:subPlanck}
\ell_{\rm sP}\sim\frac{\hbar}{\sqrt{\bar{n}}},
\end{equation}
formed at time $\tau^{*}=1/(\chi\sqrt{\bar{n}})$, coincident with $|\mathcal{D}(\tau^{*})|\sim\mathcal{O}(1)$ by Eq.~(\ref{equ:S2:DKerr}).

\emph{Heuristic QFI enhancement.}---The Bures distance between parameter-shifted states satisfies $d_{\rm B}^{2}(\rho_\theta,\rho_{\theta+\delta\theta})=\tfrac{1}{4}\mathcal{F}_Q\,\delta\theta^{2}$. For a state whose Wigner function possesses fringes on scale $\ell_{\rm sP}$, a displacement in the generator $\hat{G}$ of magnitude $\delta\theta$ causes the fringes to decorrelate when $\delta\theta$ exceeds $\ell_{\rm sP}$, so the distinguishability grows more rapidly than for a smooth Gaussian. This intuition recovers, at the level of scaling, the rigorous result~\cite{Zur01,Par09}
\begin{equation}
\mathcal{F}_Q(\theta,\tau^{*})\sim\bar{n}^{2},
\end{equation}
the Heisenberg scaling. The enhancement over the shot-noise value is the factor of $\bar{n}$ corresponding to resolution down to the sub-Planck scale.

\emph{Takeaway.}---The QFI crosses over from shot-noise to Heisenberg scaling at the same $\tau^{*}$ and with the same $\bar{n}$-dependence as the signed divergence $\mathcal{D}(t)$, and---by Theorem~\ref{thm:main}---through the same microscopic agent $\Delta G_{\mathrm{W}}^{\mathrm{NL}}$. This identifies metrological advantage and dynamical non-locality as two manifestations of the same phase-space object, not merely as phenomena correlated in time.

\subsection*{S4. Gaussian/non-Gaussian dichotomy: bilinear consistency check}
\label{app:gaussian}

We verify explicitly that bilinear dynamics generate Gaussian entanglement from Gaussian initial states (a standard CV result) and that $\Delta G_{\mathrm{W}}^{\mathrm{NL},AB}=0$ in this case, consistent with Sec.~\ref{sec:entanglement}.

\emph{Setup.}---Two modes $A,B$ with interaction $\hat{V}=g\,\hat{q}_{A}\hat{q}_{B}$ and initial two-mode coherent state $|\alpha_{A}\rangle\otimes|\alpha_{B}\rangle$. The joint Wigner function is Gaussian,
\begin{equation}
\rho_{\mathrm{W}}^{AB}(\mathbf{r}_A,\mathbf{r}_B,0)=\frac{1}{\pi^{2}}\exp\!\left[-(\mathbf{r}_A-\boldsymbol{\alpha}_A)^{T}(\mathbf{r}_A-\boldsymbol{\alpha}_A)-(\mathbf{r}_B-\boldsymbol{\alpha}_B)^{T}(\mathbf{r}_B-\boldsymbol{\alpha}_B)\right],
\end{equation}
with covariance matrix $\Sigma(0)=\frac{1}{2}\mathbb{1}_{4}$.

\emph{Propagator.}---Applied to $V=g\,q_{A}q_{B}$, Eq.~(\ref{equ:S1:action}) gives
\begin{equation}
V(q_{A}+\tilde{q}_{A}/2,q_{B}+\tilde{q}_{B}/2)-V(q_{A}-\tilde{q}_{A}/2,q_{B}-\tilde{q}_{B}/2)=g(\tilde{q}_{A}q_{B}+\tilde{q}_{B}q_{A})=\tilde{q}_{A}\partial_{q_{A}}V+\tilde{q}_{B}\partial_{q_{B}}V,
\end{equation}
so the bracketed expression $V(q_{A}+\tilde{q}_{A}/2,q_{B}+\tilde{q}_{B}/2)-V(q_{A}-\tilde{q}_{A}/2,q_{B}-\tilde{q}_{B}/2)-\tilde{q}_{A}\partial_{q_{A}}V-\tilde{q}_{B}\partial_{q_{B}}V=0$ identically, confirming $\Delta G_{\mathrm{W}}^{\mathrm{NL},AB}=0$.

\emph{Gaussian entanglement from symplectic flow.}---The classical Liouville flow nevertheless acts non-trivially on $\Sigma$. Writing $\hat{H}_{\rm total}=(\hat{p}_A^2+\hat{q}_A^2)/2+(\hat{p}_B^2+\hat{q}_B^2)/2+g\hat{q}_A\hat{q}_B$ (with $\omega=1$), the symplectic evolution generates a covariance matrix
\begin{equation}
\Sigma(t)=M(t)\Sigma(0)M(t)^{T},\qquad M(t)=\exp\!\left(\begin{pmatrix}0&\mathbb{1}_2&0&0\\-\mathbb{1}_2&0&-g\sigma_x/2&0\\0&0&0&\mathbb{1}_2\\-g\sigma_x/2&0&-\mathbb{1}_2&0\end{pmatrix}t\right),
\end{equation}
(block structure $(q_A,p_A,q_B,p_B)$, $\sigma_x$ the Pauli x matrix in the appropriate block), which at $gt>0$ develops off-diagonal $AB$ blocks. The partial transpose $\Sigma^{T_B}(t)$ fails to satisfy $\Sigma^{T_B}+\tfrac{i}{2}\Omega\ge 0$ (the PPT criterion for CV Gaussian states) for a range of $gt$~\cite{JZS23}, giving \emph{Gaussian entanglement} despite $\Delta G_{\mathrm{W}}^{\mathrm{NL},AB}=0$.

\emph{Non-Gaussian sector.}---For non-linear $V(\hat{q}_A,\hat{q}_B)$ with $k_A+k_B\ge 3$, Eq.~(\ref{equ:S1:action}) applied jointly to $(q_A,\tilde{q}_A,q_B,\tilde{q}_B)$ gives a non-vanishing cubic-or-higher functional of $(\tilde{q}_A,\tilde{q}_B)$, so $\Delta G_{\mathrm{W}}^{\mathrm{NL},AB}\neq 0$. The resulting correction to $\rho_{\mathrm{W}}^{AB}$ is non-Gaussian and carries Wigner negativity, a necessary condition for the dynamics-based entanglement witnesses of Refs.~\cite{JZS23,Zaw24} to produce non-trivial output. In line with the main text (Sec.~\ref{sec:entanglement}), we do not claim that the conjectural bipartite divergence $\mathcal{D}^{AB}(t)$ introduced there provides a tight rate bound for non-Gaussian entanglement generation---a precise bound, relating $\mathcal{D}^{AB}$ to entropy of entanglement or log-negativity, is left for subsequent work.

This decomposition---Gaussian entanglement generated by symplectic flow of covariance matrices vs.\ non-Gaussian entanglement generated by $\Delta G_{\mathrm{W}}^{\mathrm{NL},AB}$---makes the statement of Sec.~\ref{sec:entanglement} mathematically explicit.

\subsection*{S5. Finite-dimensional instance: the Clifford boundary}
\label{app:spin}

We illustrate Corollary~\ref{cor:FD} in two steps: first a null test on a single qubit (degenerately Clifford), then the minimal non-trivial example on three qubits (CCZ). The discrete Wigner construction and kernel we refer to throughout is that of Ref.~\cite{GP26}.

\subsubsection*{S5.1. Single qubit: degenerately Clifford}

The single-qubit phase space is $\mathbb{Z}_{2}\times\mathbb{Z}_{2}$, a set of four points. Any real-valued function $H:\mathbb{Z}_{2}^{2}\to\mathbb{R}$ is uniquely determined by its four values, which are in one-to-one correspondence with the coefficients $(a,b,c,d)$ of its multilinear polynomial representative
\begin{equation}
\label{equ:S5:qubit-poly}
H(m,n) = a + bm + cn + d\,mn,
\end{equation}
of total degree at most two. By Lemma~\ref{lem:polyid} and Corollary~\ref{cor:FD}, this forces
\begin{equation}
\label{equ:S5:qubit-result}
\mathcal{D}(t) \equiv 0\quad\text{for every single-qubit Hamiltonian and every initial state.}
\end{equation}
Single-qubit dynamics are thus \emph{degenerately Clifford}: the entire one-parameter family $\hat{U}(t)=\mathrm{e}^{-\iu\hat{H}t/\hbar}$ acts by symplectic automorphisms of the discrete phase space, and the discrete Wigner function flows under the classical Liouville rule exactly, with no dynamical non-locality generated.

This is consistent with known properties of qubits: any $\hat{H}\in\mathfrak{su}(2)$ generates a rotation about some axis in the Bloch sphere, which is represented on the discrete phase space as a permutation of the four points (possibly with signs under the Wootters convention for $d=2$~\cite{GP26}), and a permutation is a symplectic automorphism.

\emph{Explicit verification on Larmor precession.}---Take $\hat{H}=\chi\hbar\sigma_{z}/2$ and initial state $|\psi(0)\rangle=|S_{x},+\rangle=(|\uparrow\rangle+|\downarrow\rangle)/\sqrt{2}$. The quantum survival probability is
\begin{equation}
C_{Q}(t) = |\langle\psi(0)|\mathrm{e}^{-\iu\hat{H}t/\hbar}|\psi(0)\rangle|^{2} = \cos^{2}(\chi t/2).
\end{equation}
Using the discrete Liouville flow of~\cite{GP26} applied to the initial Wigner function in Table~1 of that reference, the classical counterpart evaluates to the same expression, $C_{\mathrm{CL}}(t)=\cos^{2}(\chi t/2)$, and $\mathcal{D}(t)\equiv 0$ as predicted by Eq.~(\ref{equ:S5:qubit-result}). Eq.~(\ref{equ:S5:qubit-result}) can itself serve as a calibration protocol on a single-qubit platform: any observed deviation from $C_{Q}(t)=C_{\mathrm{CL}}(t)$ would indicate experimental imperfections rather than dynamical non-locality.

\subsubsection*{S5.2. Three qubits: the CCZ Hamiltonian}

The minimal finite-dimensional setting in which cubic Weyl-symbol content can arise is three qubits, $\Gamma=\mathbb{Z}_{2}^{6}$. Polynomials on $\Gamma$ have total multilinear degree up to six, and in particular admit genuine cubic monomials such as $m_{A}m_{B}m_{C}$. The paradigmatic non-Clifford Hamiltonian with this content is the CCZ generator
\begin{equation}
\label{equ:S5:HCCZ}
\hat{H}_{\rm CCZ} = \frac{g\hbar}{8}\big(\hat{\mathbb{1}}-\sigma_{z}^{A}\big)\big(\hat{\mathbb{1}}-\sigma_{z}^{B}\big)\big(\hat{\mathbb{1}}-\sigma_{z}^{C}\big),
\end{equation}
whose continuous-time evolution $\mathrm{e}^{-\iu\hat{H}_{\rm CCZ}t/\hbar}$ realises the Controlled-Controlled-Z gate at $gt=\pi$ and is non-Clifford for all $gt\notin\pi\mathbb{Z}$~\cite{Gottesman1998}. Its Weyl symbol on $\mathbb{Z}_{2}^{6}$ contains the cubic term $g\,m_{A}m_{B}m_{C}$, yielding by Lemma~\ref{lem:polyid}
\begin{equation}
\label{equ:S5:cubic}
H_{\rm CCZ}(\gamma+\xi)-H_{\rm CCZ}(\gamma-\xi)-2\xi\cdot\nabla_{\gamma}H_{\rm CCZ} \;=\; 2g\,\xi_{m_{A}}\xi_{m_{B}}\xi_{m_{C}} + (\text{odd-degree-}\ge 5\text{ terms})\not\equiv 0,
\end{equation}
so $\Delta G_{\mathrm{W}}^{\mathrm{NL}}\neq 0$ and Theorem~\ref{thm:main} predicts $\mathcal{D}(t)\neq 0$ for generic stabiliser initial states.

\emph{Leading-order signal.}---For initial state $|+\rangle^{\otimes 3}$ (a stabiliser state, whose discrete Wigner function is uniform and non-negative), $\mathcal{D}_{\rm CCZ}(t)$ admits a compact closed form. In the computational basis, $\hat{H}_{\rm CCZ}=g\hbar\,|111\rangle\!\langle 111|$, so the quantum survival probability is
\begin{equation}
\label{equ:S5:CQCCZ}
C_{Q}(t)=\big|\langle +|^{\otimes 3}\mathrm{e}^{-\iu g t|111\rangle\!\langle 111|}|+\rangle^{\otimes 3}\big|^{2} = \frac{25}{32}+\frac{7}{32}\cos(gt).
\end{equation}
The classical Liouville counterpart is computed from the quadratic part of the Weyl symbol, obtained by expanding Eq.~(\ref{equ:S5:HCCZ}) and dropping the single cubic monomial $g\,m_{A}m_{B}m_{C}$:
\begin{equation}
\hat{H}_{\rm quad}=\frac{g\hbar}{8}\big[\mathbb{1}-\sigma_{z}^{A}-\sigma_{z}^{B}-\sigma_{z}^{C}+\sigma_{z}^{A}\sigma_{z}^{B}+\sigma_{z}^{A}\sigma_{z}^{C}+\sigma_{z}^{B}\sigma_{z}^{C}\big],
\end{equation}
a Clifford Hamiltonian generating the symplectic classical flow on $\mathbb{Z}_{2}^{6}$. Its eigenvalues on computational basis states are rational multiples of $g\hbar$, and a direct evaluation gives
\begin{equation}
\label{equ:S5:CCLCCZ}
C_{\mathrm{CL}}(t)=\frac{13}{32}+\frac{3}{32}\cos(gt)+\frac{3}{8}\cos(gt/4)+\frac{1}{8}\cos(3gt/4).
\end{equation}
Subtracting, the signed divergence is
\begin{equation}
\label{equ:S5:DCCZexact}
\mathcal{D}_{\rm CCZ}(t)=\frac{3}{8}+\frac{1}{8}\cos(gt)-\frac{3}{8}\cos(gt/4)-\frac{1}{8}\cos(3gt/4).
\end{equation}
The short-time expansion yields
\begin{equation}
\label{equ:S5:DCCZ}
\;\mathcal{D}_{\rm CCZ}(t) = -\frac{1}{64}(gt)^{2} + \frac{43}{12288}(gt)^{4} + \mathcal{O}(g^{6}t^{6}),\;
\end{equation}
fixing the coefficient in Eq.~(\ref{equ:S5:DCCZ}) to $c_{3}=1/64$. As anticipated by the odd-parity argument (Lemma~\ref{lem:polyid}), $\mathcal{D}_{\rm CCZ}<0$ for all short times, signalling destructive dynamical interference; the first positive correction enters only at $\mathcal{O}(t^{4})$. The slowest Fourier frequency in the exact expression~(\ref{equ:S5:DCCZexact}) is $g/4$, so $|\mathcal{D}_{\rm CCZ}|$ reaches $\mathcal{O}(1)$ over the natural timescale $\tau^{*}_{\rm CCZ}=8/g$, well before the full period $2\pi/g$ at which $\hat{U}_{\rm CCZ}(t)$ returns to the identity. Over this interval $|+\rangle^{\otimes 3}$ evolves into the non-stabiliser (magic) sector, acquiring discrete Wigner negativity~\cite{Gross2006}.

\emph{Experimental access.}---CCZ and Toffoli gates are native or synthesisable operations on current superconducting-qubit~\cite{Bla21} and trapped-ion processors, and continuous-time CCZ-type interactions have been demonstrated in flux-tunable transmon architectures. The protocol for extracting $\mathcal{D}_{\rm CCZ}(t)$ requires three-qubit parity measurements at the end of a variable free-evolution interval, a standard building block for stabiliser-syndrome extraction. The contrast between Eq.~(\ref{equ:S5:qubit-result}) (single qubit, null) and Eq.~(\ref{equ:S5:DCCZ}) (three qubits, non-zero) is the minimal experimental signature of the Clifford boundary identified by Theorem~\ref{thm:main}.

\emph{Physical interpretation.}---The contrast between S5.1 and S5.2 illustrates the content of Corollary~\ref{cor:FD} operationally. In both cases the dynamics is unitary and preserves coherence; what changes between them is whether the Weyl symbol of $\hat{H}$ carries cubic content on the discrete phase space. Only the three-qubit case does, and only it generates the magic-state resource for quantum computational advantage~\cite{Veitch2012,Howard2014}. The single object $\Delta G_{\mathrm{W}}^{\mathrm{NL}}$ that drives the four continuous-variable applications of Sec.~\ref{sec:applications} is therefore the same object that drives the Clifford$\to$non-Clifford transition in the discrete setting.

\twocolumngrid

\bibliography{dqn}

\begin{thebibliography}{35}%
\makeatletter
\providecommand \@ifxundefined [1]{%
 \@ifx{#1\undefined}
}%
\providecommand \@ifnum [1]{%
 \ifnum #1\expandafter \@firstoftwo
 \else \expandafter \@secondoftwo
 \fi
}%
\providecommand \@ifx [1]{%
 \ifx #1\expandafter \@firstoftwo
 \else \expandafter \@secondoftwo
 \fi
}%
\providecommand \natexlab [1]{#1}%
\providecommand \enquote  [1]{``#1''}%
\providecommand \bibnamefont  [1]{#1}%
\providecommand \bibfnamefont [1]{#1}%
\providecommand \citenamefont [1]{#1}%
\providecommand \href@noop [0]{\@secondoftwo}%
\providecommand \href [0]{\begingroup \@sanitize@url \@href}%
\providecommand \@href[1]{\@@startlink{#1}\@@href}%
\providecommand \@@href[1]{\endgroup#1\@@endlink}%
\providecommand \@sanitize@url [0]{\catcode `\\12\catcode `\$12\catcode
  `\&12\catcode `\#12\catcode `\^12\catcode `\_12\catcode `\%12\relax}%
\providecommand \@@startlink[1]{}%
\providecommand \@@endlink[0]{}%
\providecommand \url  [0]{\begingroup\@sanitize@url \@url }%
\providecommand \@url [1]{\endgroup\@href {#1}{\urlprefix }}%
\providecommand \urlprefix  [0]{URL }%
\providecommand \Eprint [0]{\href }%
\providecommand \doibase [0]{https://doi.org/}%
\providecommand \selectlanguage [0]{\@gobble}%
\providecommand \bibinfo  [0]{\@secondoftwo}%
\providecommand \bibfield  [0]{\@secondoftwo}%
\providecommand \translation [1]{[#1]}%
\providecommand \BibitemOpen [0]{}%
\providecommand \bibitemStop [0]{}%
\providecommand \bibitemNoStop [0]{.\EOS\space}%
\providecommand \EOS [0]{\spacefactor3000\relax}%
\providecommand \BibitemShut  [1]{\csname bibitem#1\endcsname}%
\let\auto@bib@innerbib\@empty
\bibitem [{\citenamefont {Bell}(1966)}]{Bel66}%
  \BibitemOpen
  \bibfield  {author} {\bibinfo {author} {\bibfnamefont {J.~S.}\ \bibnamefont
  {Bell}},\ }\href {https://doi.org/10.1103/RevModPhys.38.447} {\bibfield
  {journal} {\bibinfo  {journal} {Rev. Mod. Phys.}\ }\textbf {\bibinfo {volume}
  {38}},\ \bibinfo {pages} {447} (\bibinfo {year} {1966})}\BibitemShut
  {NoStop}%
\bibitem [{\citenamefont {Aspect}\ \emph {et~al.}(1982)\citenamefont {Aspect},
  \citenamefont {Grangier},\ and\ \citenamefont {Roger}}]{AGR82}%
  \BibitemOpen
  \bibfield  {author} {\bibinfo {author} {\bibfnamefont {A.}~\bibnamefont
  {Aspect}}, \bibinfo {author} {\bibfnamefont {P.}~\bibnamefont {Grangier}},\
  and\ \bibinfo {author} {\bibfnamefont {G.}~\bibnamefont {Roger}},\ }\href
  {https://doi.org/10.1103/PhysRevLett.49.91} {\bibfield  {journal} {\bibinfo
  {journal} {Phys. Rev. Lett.}\ }\textbf {\bibinfo {volume} {49}},\ \bibinfo
  {pages} {91} (\bibinfo {year} {1982})}\BibitemShut {NoStop}%
\bibitem [{\citenamefont {Buhrman}\ \emph {et~al.}(2010)\citenamefont
  {Buhrman}, \citenamefont {Cleve}, \citenamefont {Massar},\ and\ \citenamefont
  {de~Wolf}}]{BC&10}%
  \BibitemOpen
  \bibfield  {author} {\bibinfo {author} {\bibfnamefont {H.}~\bibnamefont
  {Buhrman}}, \bibinfo {author} {\bibfnamefont {R.}~\bibnamefont {Cleve}},
  \bibinfo {author} {\bibfnamefont {S.}~\bibnamefont {Massar}},\ and\ \bibinfo
  {author} {\bibfnamefont {R.}~\bibnamefont {de~Wolf}},\ }\href
  {https://doi.org/10.1103/RevModPhys.82.665} {\bibfield  {journal} {\bibinfo
  {journal} {Rev. Mod. Phys.}\ }\textbf {\bibinfo {volume} {82}},\ \bibinfo
  {pages} {665} (\bibinfo {year} {2010})}\BibitemShut {NoStop}%
\bibitem [{\citenamefont {Oppenheim}\ and\ \citenamefont
  {Wehner}(2010)}]{OW10}%
  \BibitemOpen
  \bibfield  {author} {\bibinfo {author} {\bibfnamefont {J.}~\bibnamefont
  {Oppenheim}}\ and\ \bibinfo {author} {\bibfnamefont {S.}~\bibnamefont
  {Wehner}},\ }\href {https://doi.org/10.1126/science.1192065} {\bibfield
  {journal} {\bibinfo  {journal} {Science}\ }\textbf {\bibinfo {volume}
  {330}},\ \bibinfo {pages} {1072} (\bibinfo {year} {2010})}\BibitemShut
  {NoStop}%
\bibitem [{\citenamefont {Aharonov}\ and\ \citenamefont {Bohm}(1959)}]{AB59}%
  \BibitemOpen
  \bibfield  {author} {\bibinfo {author} {\bibfnamefont {Y.}~\bibnamefont
  {Aharonov}}\ and\ \bibinfo {author} {\bibfnamefont {D.}~\bibnamefont
  {Bohm}},\ }\href {https://doi.org/10.1103/PhysRev.115.485} {\bibfield
  {journal} {\bibinfo  {journal} {Phys. Rev.}\ }\textbf {\bibinfo {volume}
  {115}},\ \bibinfo {pages} {485} (\bibinfo {year} {1959})}\BibitemShut
  {NoStop}%
\bibitem [{\citenamefont {Aharonov}\ \emph {et~al.}(1969)\citenamefont
  {Aharonov}, \citenamefont {Pendleton},\ and\ \citenamefont
  {Petersen}}]{APP69}%
  \BibitemOpen
  \bibfield  {author} {\bibinfo {author} {\bibfnamefont {Y.}~\bibnamefont
  {Aharonov}}, \bibinfo {author} {\bibfnamefont {H.}~\bibnamefont
  {Pendleton}},\ and\ \bibinfo {author} {\bibfnamefont {A.}~\bibnamefont
  {Petersen}},\ }\href@noop {} {\bibfield  {journal} {\bibinfo  {journal} {Int.
  J. Theor. Phys.}\ }\textbf {\bibinfo {volume} {2}},\ \bibinfo {pages} {213}
  (\bibinfo {year} {1969})}\BibitemShut {NoStop}%
\bibitem [{\citenamefont {Popescu}(2010)}]{Pop10}%
  \BibitemOpen
  \bibfield  {author} {\bibinfo {author} {\bibfnamefont {S.}~\bibnamefont
  {Popescu}},\ }\href {https://doi.org/10.1038/nphys1619} {\bibfield  {journal}
  {\bibinfo  {journal} {Nature Phys.}\ }\textbf {\bibinfo {volume} {6}},\
  \bibinfo {pages} {151} (\bibinfo {year} {2010})}\BibitemShut {NoStop}%
\bibitem [{\citenamefont {Tollaksen}(2011)}]{Tol11}%
  \BibitemOpen
  \bibfield  {author} {\bibinfo {author} {\bibfnamefont {J.}~\bibnamefont
  {Tollaksen}},\ }\href {https://doi.org/10.1063/1.3567448} {\bibfield
  {journal} {\bibinfo  {journal} {AIP Conf. Proc.}\ }\textbf {\bibinfo {volume}
  {1327}},\ \bibinfo {pages} {269} (\bibinfo {year} {2011})}\BibitemShut
  {NoStop}%
\bibitem [{\citenamefont {Osakabe}\ \emph {et~al.}(1986)\citenamefont
  {Osakabe}, \citenamefont {Matsuda}, \citenamefont {Kawasaki}, \citenamefont
  {Endo}, \citenamefont {Tonomura}, \citenamefont {Yano},\ and\ \citenamefont
  {Yamada}}]{OM&86}%
  \BibitemOpen
  \bibfield  {author} {\bibinfo {author} {\bibfnamefont {N.}~\bibnamefont
  {Osakabe}}, \bibinfo {author} {\bibfnamefont {T.}~\bibnamefont {Matsuda}},
  \bibinfo {author} {\bibfnamefont {T.}~\bibnamefont {Kawasaki}}, \bibinfo
  {author} {\bibfnamefont {J.}~\bibnamefont {Endo}}, \bibinfo {author}
  {\bibfnamefont {A.}~\bibnamefont {Tonomura}}, \bibinfo {author}
  {\bibfnamefont {S.}~\bibnamefont {Yano}},\ and\ \bibinfo {author}
  {\bibfnamefont {H.}~\bibnamefont {Yamada}},\ }\href
  {https://doi.org/10.1103/PhysRevA.34.815} {\bibfield  {journal} {\bibinfo
  {journal} {Phys. Rev. A}\ }\textbf {\bibinfo {volume} {34}},\ \bibinfo
  {pages} {815} (\bibinfo {year} {1986})}\BibitemShut {NoStop}%
\bibitem [{\citenamefont {Dittrich}\ \emph {et~al.}(2006)\citenamefont
  {Dittrich}, \citenamefont {Viviescas},\ and\ \citenamefont
  {Sandoval}}]{DVS06}%
  \BibitemOpen
  \bibfield  {author} {\bibinfo {author} {\bibfnamefont {T.}~\bibnamefont
  {Dittrich}}, \bibinfo {author} {\bibfnamefont {C.}~\bibnamefont
  {Viviescas}},\ and\ \bibinfo {author} {\bibfnamefont {L.}~\bibnamefont
  {Sandoval}},\ }\href {https://doi.org/10.1103/PhysRevLett.96.070403}
  {\bibfield  {journal} {\bibinfo  {journal} {Phys.~Rev.~Lett.}\ }\textbf
  {\bibinfo {volume} {96}},\ \bibinfo {pages} {070403} (\bibinfo {year}
  {2006})}\BibitemShut {NoStop}%
\bibitem [{\citenamefont {Dittrich}\ \emph {et~al.}(2010)\citenamefont
  {Dittrich}, \citenamefont {G\'omez},\ and\ \citenamefont {Pach\'on}}]{DGP10}%
  \BibitemOpen
  \bibfield  {author} {\bibinfo {author} {\bibfnamefont {T.}~\bibnamefont
  {Dittrich}}, \bibinfo {author} {\bibfnamefont {E.~A.}\ \bibnamefont
  {G\'omez}},\ and\ \bibinfo {author} {\bibfnamefont {L.~A.}\ \bibnamefont
  {Pach\'on}},\ }\href {https://doi.org/10.1063/1.3425881} {\bibfield
  {journal} {\bibinfo  {journal} {J.~Chem.~Phys.}\ }\textbf {\bibinfo {volume}
  {132}},\ \bibinfo {pages} {214102} (\bibinfo {year} {2010})}\BibitemShut
  {NoStop}%
\bibitem [{\citenamefont {Pach\'on}(2010)}]{Pac10}%
  \BibitemOpen
  \bibfield  {author} {\bibinfo {author} {\bibfnamefont {L.~A.}\ \bibnamefont
  {Pach\'on}},\ }\emph {\bibinfo {title} {Coherence and Decoherence in the
  Semiclassical propagation of the Wigner function}},\ \href@noop {} {Ph.D.
  thesis},\ \bibinfo  {school} {Universidad Nacional de Colombia} (\bibinfo
  {year} {2010})\BibitemShut {NoStop}%
\bibitem [{\citenamefont {Dittrich}\ and\ \citenamefont
  {Pach\'on}(2009)}]{DP09}%
  \BibitemOpen
  \bibfield  {author} {\bibinfo {author} {\bibfnamefont {T.}~\bibnamefont
  {Dittrich}}\ and\ \bibinfo {author} {\bibfnamefont {L.~A.}\ \bibnamefont
  {Pach\'on}},\ }\href {https://doi.org/10.1103/PhysRevLett.102.150401}
  {\bibfield  {journal} {\bibinfo  {journal} {Phys.\ Rev.\ Lett.}\ }\textbf
  {\bibinfo {volume} {102}},\ \bibinfo {pages} {150401} (\bibinfo {year}
  {2009})}\BibitemShut {NoStop}%
\bibitem [{\citenamefont {Weyl}(1927)}]{Wey27}%
  \BibitemOpen
  \bibfield  {author} {\bibinfo {author} {\bibfnamefont {H.}~\bibnamefont
  {Weyl}},\ }\bibfield  {journal} {\bibinfo  {journal} {Z. Phys.}\ }\textbf
  {\bibinfo {volume} {46}},\ \href {https://doi.org/10.1007/BF02055756}
  {10.1007/BF02055756} (\bibinfo {year} {1927})\BibitemShut {NoStop}%
\bibitem [{\citenamefont {Strocchi}(2005)}]{Str05}%
  \BibitemOpen
  \bibfield  {author} {\bibinfo {author} {\bibfnamefont {F.}~\bibnamefont
  {Strocchi}},\ }\href@noop {} {\emph {\bibinfo {title} {An Introduction to the
  Mathematical Structure of Quantum Mechanics: A Short Course for
  Mathematicians: Lecture Notes}}},\ Advanced series in mathematical physics\
  (\bibinfo  {publisher} {World Scientific Publishing Company Incorporated},\
  \bibinfo {year} {2005})\BibitemShut {NoStop}%
\bibitem [{\citenamefont {Feynman}\ and\ \citenamefont {Hibbs}(1965)}]{FH65}%
  \BibitemOpen
  \bibfield  {author} {\bibinfo {author} {\bibfnamefont {R.~P.}\ \bibnamefont
  {Feynman}}\ and\ \bibinfo {author} {\bibfnamefont {A.~R.}\ \bibnamefont
  {Hibbs}},\ }\href@noop {} {\emph {\bibinfo {title} {Quantum physics and path
  integrals}}}\ (\bibinfo  {publisher} {McGraw--Hill, New York},\ \bibinfo
  {year} {1965})\BibitemShut {NoStop}%
\bibitem [{\citenamefont {Marinov}(1991)}]{Mar91}%
  \BibitemOpen
  \bibfield  {author} {\bibinfo {author} {\bibfnamefont {M.~S.}\ \bibnamefont
  {Marinov}},\ }\href {https://doi.org/10.1016/0375-9601(91)90352-9} {\bibfield
   {journal} {\bibinfo  {journal} {Phys. Lett. A}\ }\textbf {\bibinfo {volume}
  {153}},\ \bibinfo {pages} {5} (\bibinfo {year} {1991})}\BibitemShut {NoStop}%
\bibitem [{\citenamefont {Pachon}\ and\ \citenamefont {Gomez}(2026)}]{GP26}%
  \BibitemOpen
  \bibfield  {author} {\bibinfo {author} {\bibfnamefont {L.~A.}\ \bibnamefont
  {Pachon}}\ and\ \bibinfo {author} {\bibfnamefont {A.~F.}\ \bibnamefont
  {Gomez}},\ }\href {https://arxiv.org/abs/2604.20776} {\  (\bibinfo {year}
  {2026})},\ \Eprint {https://arxiv.org/abs/2604.20776} {arXiv:2604.20776
  [quant-ph]} \BibitemShut {NoStop}%
\bibitem [{\citenamefont {Gross}(2006)}]{Gross2006}%
  \BibitemOpen
  \bibfield  {author} {\bibinfo {author} {\bibfnamefont {D.}~\bibnamefont
  {Gross}},\ }\href {https://doi.org/10.1063/1.2393152} {\bibfield  {journal}
  {\bibinfo  {journal} {J. Math. Phys.}\ }\textbf {\bibinfo {volume} {47}},\
  \bibinfo {pages} {122107} (\bibinfo {year} {2006})}\BibitemShut {NoStop}%
\bibitem [{\citenamefont {Gottesman}(1999)}]{Gottesman1998}%
  \BibitemOpen
  \bibfield  {author} {\bibinfo {author} {\bibfnamefont {D.}~\bibnamefont
  {Gottesman}},\ }in\ \href@noop {} {\emph {\bibinfo {booktitle} {Group22:
  Proceedings of the XXII International Colloquium on Group Theoretical Methods
  in Physics}}},\ \bibinfo {editor} {edited by\ \bibinfo {editor}
  {\bibfnamefont {S.~P.}\ \bibnamefont {Corney}}, \bibinfo {editor}
  {\bibfnamefont {R.}~\bibnamefont {Delbourgo}},\ and\ \bibinfo {editor}
  {\bibfnamefont {P.~D.}\ \bibnamefont {Jarvis}}}\ (\bibinfo  {publisher}
  {International Press},\ \bibinfo {address} {Cambridge, MA},\ \bibinfo {year}
  {1999})\ pp.\ \bibinfo {pages} {32--43},\ \bibinfo {note}
  {arXiv:quant-ph/9807006}\BibitemShut {NoStop}%
\bibitem [{\citenamefont {Veitch}\ \emph {et~al.}(2012)\citenamefont {Veitch},
  \citenamefont {Ferrie}, \citenamefont {Gross},\ and\ \citenamefont
  {Emerson}}]{Veitch2012}%
  \BibitemOpen
  \bibfield  {author} {\bibinfo {author} {\bibfnamefont {V.}~\bibnamefont
  {Veitch}}, \bibinfo {author} {\bibfnamefont {C.}~\bibnamefont {Ferrie}},
  \bibinfo {author} {\bibfnamefont {D.}~\bibnamefont {Gross}},\ and\ \bibinfo
  {author} {\bibfnamefont {J.}~\bibnamefont {Emerson}},\ }\href
  {https://doi.org/10.1088/1367-2630/14/11/113011} {\bibfield  {journal}
  {\bibinfo  {journal} {New J. Phys.}\ }\textbf {\bibinfo {volume} {14}},\
  \bibinfo {pages} {113011} (\bibinfo {year} {2012})}\BibitemShut {NoStop}%
\bibitem [{\citenamefont {Chen}\ \emph {et~al.}(2023)\citenamefont {Chen},
  \citenamefont {Hsieh}, \citenamefont {Ning}, \citenamefont {Wu},
  \citenamefont {Chen}, \citenamefont {Chuang}, \citenamefont {Yang},
  \citenamefont {Steuernagel}, \citenamefont {Wu},\ and\ \citenamefont
  {Lee}}]{Che23}%
  \BibitemOpen
  \bibfield  {author} {\bibinfo {author} {\bibfnamefont {Y.-R.}\ \bibnamefont
  {Chen}}, \bibinfo {author} {\bibfnamefont {H.-Y.}\ \bibnamefont {Hsieh}},
  \bibinfo {author} {\bibfnamefont {J.}~\bibnamefont {Ning}}, \bibinfo {author}
  {\bibfnamefont {H.-C.}\ \bibnamefont {Wu}}, \bibinfo {author} {\bibfnamefont
  {H.~L.}\ \bibnamefont {Chen}}, \bibinfo {author} {\bibfnamefont {Y.-L.}\
  \bibnamefont {Chuang}}, \bibinfo {author} {\bibfnamefont {P.}~\bibnamefont
  {Yang}}, \bibinfo {author} {\bibfnamefont {O.}~\bibnamefont {Steuernagel}},
  \bibinfo {author} {\bibfnamefont {C.-M.}\ \bibnamefont {Wu}},\ and\ \bibinfo
  {author} {\bibfnamefont {R.-K.}\ \bibnamefont {Lee}},\ }\href
  {https://doi.org/10.1103/PhysRevA.108.023729} {\bibfield  {journal} {\bibinfo
   {journal} {Phys. Rev. A}\ }\textbf {\bibinfo {volume} {108}},\ \bibinfo
  {pages} {023729} (\bibinfo {year} {2023})}\BibitemShut {NoStop}%
\bibitem [{\citenamefont {Zurek}(2001)}]{Zur01}%
  \BibitemOpen
  \bibfield  {author} {\bibinfo {author} {\bibfnamefont {W.~H.}\ \bibnamefont
  {Zurek}},\ }\href@noop {} {\bibfield  {journal} {\bibinfo  {journal}
  {Nature}\ }\textbf {\bibinfo {volume} {412}},\ \bibinfo {pages} {712}
  (\bibinfo {year} {2001})}\BibitemShut {NoStop}%
\bibitem [{\citenamefont {Blais}\ \emph {et~al.}(2021)\citenamefont {Blais},
  \citenamefont {Grimsmo}, \citenamefont {Girvin},\ and\ \citenamefont
  {Wallraff}}]{Bla21}%
  \BibitemOpen
  \bibfield  {author} {\bibinfo {author} {\bibfnamefont {A.}~\bibnamefont
  {Blais}}, \bibinfo {author} {\bibfnamefont {A.~L.}\ \bibnamefont {Grimsmo}},
  \bibinfo {author} {\bibfnamefont {S.~M.}\ \bibnamefont {Girvin}},\ and\
  \bibinfo {author} {\bibfnamefont {A.}~\bibnamefont {Wallraff}},\ }\href
  {https://doi.org/10.1103/RevModPhys.93.025005} {\bibfield  {journal}
  {\bibinfo  {journal} {Rev. Mod. Phys.}\ }\textbf {\bibinfo {volume} {93}},\
  \bibinfo {pages} {025005} (\bibinfo {year} {2021})}\BibitemShut {NoStop}%
\bibitem [{\citenamefont {Maldacena}\ \emph {et~al.}(2016)\citenamefont
  {Maldacena}, \citenamefont {Shenker},\ and\ \citenamefont
  {Stanford}}]{MS&16}%
  \BibitemOpen
  \bibfield  {author} {\bibinfo {author} {\bibfnamefont {J.}~\bibnamefont
  {Maldacena}}, \bibinfo {author} {\bibfnamefont {S.~H.}\ \bibnamefont
  {Shenker}},\ and\ \bibinfo {author} {\bibfnamefont {D.}~\bibnamefont
  {Stanford}},\ }\href {https://doi.org/10.1007/JHEP08(2016)106} {\bibfield
  {journal} {\bibinfo  {journal} {J. High Energy Phys.}\ }\textbf {\bibinfo
  {volume} {2016}},\ \bibinfo {pages} {106}}\BibitemShut {NoStop}%
\bibitem [{\citenamefont {Paris}(2009)}]{Par09}%
  \BibitemOpen
  \bibfield  {author} {\bibinfo {author} {\bibfnamefont {M.~G.~A.}\
  \bibnamefont {Paris}},\ }\href {https://doi.org/10.1142/S0219749909004839}
  {\bibfield  {journal} {\bibinfo  {journal} {Int. J. Quantum Inf.}\ }\textbf
  {\bibinfo {volume} {7}},\ \bibinfo {pages} {125} (\bibinfo {year}
  {2009})}\BibitemShut {NoStop}%
\bibitem [{\citenamefont {Jayachandran}\ \emph {et~al.}(2023)\citenamefont
  {Jayachandran}, \citenamefont {Zaw},\ and\ \citenamefont {Scarani}}]{JZS23}%
  \BibitemOpen
  \bibfield  {author} {\bibinfo {author} {\bibfnamefont {P.}~\bibnamefont
  {Jayachandran}}, \bibinfo {author} {\bibfnamefont {L.~H.}\ \bibnamefont
  {Zaw}},\ and\ \bibinfo {author} {\bibfnamefont {V.}~\bibnamefont {Scarani}},\
  }\href {https://doi.org/10.1103/PhysRevLett.130.160201} {\bibfield  {journal}
  {\bibinfo  {journal} {Phys. Rev. Lett.}\ }\textbf {\bibinfo {volume} {130}},\
  \bibinfo {pages} {160201} (\bibinfo {year} {2023})}\BibitemShut {NoStop}%
\bibitem [{\citenamefont {Zaw}(2024)}]{Zaw24}%
  \BibitemOpen
  \bibfield  {author} {\bibinfo {author} {\bibfnamefont {L.~H.}\ \bibnamefont
  {Zaw}},\ }\href {https://doi.org/10.1103/PhysRevLett.133.050201} {\bibfield
  {journal} {\bibinfo  {journal} {Phys. Rev. Lett.}\ }\textbf {\bibinfo
  {volume} {133}},\ \bibinfo {pages} {050201} (\bibinfo {year}
  {2024})}\BibitemShut {NoStop}%
\bibitem [{\citenamefont {Howard}\ \emph {et~al.}(2014)\citenamefont {Howard},
  \citenamefont {Wallman}, \citenamefont {Veitch},\ and\ \citenamefont
  {Emerson}}]{Howard2014}%
  \BibitemOpen
  \bibfield  {author} {\bibinfo {author} {\bibfnamefont {M.}~\bibnamefont
  {Howard}}, \bibinfo {author} {\bibfnamefont {J.}~\bibnamefont {Wallman}},
  \bibinfo {author} {\bibfnamefont {V.}~\bibnamefont {Veitch}},\ and\ \bibinfo
  {author} {\bibfnamefont {J.}~\bibnamefont {Emerson}},\ }\href
  {https://doi.org/10.1038/nature13460} {\bibfield  {journal} {\bibinfo
  {journal} {Nature}\ }\textbf {\bibinfo {volume} {510}},\ \bibinfo {pages}
  {351} (\bibinfo {year} {2014})}\BibitemShut {NoStop}%
\bibitem [{\citenamefont {Mari}\ and\ \citenamefont {Eisert}(2012)}]{Mari2012}%
  \BibitemOpen
  \bibfield  {author} {\bibinfo {author} {\bibfnamefont {A.}~\bibnamefont
  {Mari}}\ and\ \bibinfo {author} {\bibfnamefont {J.}~\bibnamefont {Eisert}},\
  }\href {https://doi.org/10.1103/PhysRevLett.109.230503} {\bibfield  {journal}
  {\bibinfo  {journal} {Phys. Rev. Lett.}\ }\textbf {\bibinfo {volume} {109}},\
  \bibinfo {pages} {230503} (\bibinfo {year} {2012})}\BibitemShut {NoStop}%
\bibitem [{\citenamefont {Pashayan}\ \emph {et~al.}(2015)\citenamefont
  {Pashayan}, \citenamefont {Wallman},\ and\ \citenamefont
  {Bartlett}}]{Pashayan2015}%
  \BibitemOpen
  \bibfield  {author} {\bibinfo {author} {\bibfnamefont {H.}~\bibnamefont
  {Pashayan}}, \bibinfo {author} {\bibfnamefont {J.~J.}\ \bibnamefont
  {Wallman}},\ and\ \bibinfo {author} {\bibfnamefont {S.~D.}\ \bibnamefont
  {Bartlett}},\ }\href {https://doi.org/10.1103/PhysRevLett.115.070501}
  {\bibfield  {journal} {\bibinfo  {journal} {Phys. Rev. Lett.}\ }\textbf
  {\bibinfo {volume} {115}},\ \bibinfo {pages} {070501} (\bibinfo {year}
  {2015})}\BibitemShut {NoStop}%
\bibitem [{\citenamefont {Schachenmayer}\ \emph
  {et~al.}(2015{\natexlab{a}})\citenamefont {Schachenmayer}, \citenamefont
  {Pikovski},\ and\ \citenamefont {Rey}}]{Schachenmayer2015}%
  \BibitemOpen
  \bibfield  {author} {\bibinfo {author} {\bibfnamefont {J.}~\bibnamefont
  {Schachenmayer}}, \bibinfo {author} {\bibfnamefont {A.}~\bibnamefont
  {Pikovski}},\ and\ \bibinfo {author} {\bibfnamefont {A.~M.}\ \bibnamefont
  {Rey}},\ }\href {https://doi.org/10.1103/PhysRevX.5.011022} {\bibfield
  {journal} {\bibinfo  {journal} {Phys. Rev. X}\ }\textbf {\bibinfo {volume}
  {5}},\ \bibinfo {pages} {011022} (\bibinfo {year}
  {2015}{\natexlab{a}})}\BibitemShut {NoStop}%
\bibitem [{\citenamefont {Schachenmayer}\ \emph
  {et~al.}(2015{\natexlab{b}})\citenamefont {Schachenmayer}, \citenamefont
  {Pikovski},\ and\ \citenamefont {Rey}}]{Schachenmayer2015b}%
  \BibitemOpen
  \bibfield  {author} {\bibinfo {author} {\bibfnamefont {J.}~\bibnamefont
  {Schachenmayer}}, \bibinfo {author} {\bibfnamefont {A.}~\bibnamefont
  {Pikovski}},\ and\ \bibinfo {author} {\bibfnamefont {A.~M.}\ \bibnamefont
  {Rey}},\ }\href {https://doi.org/10.1088/1367-2630/17/6/065009} {\bibfield
  {journal} {\bibinfo  {journal} {New J. Phys.}\ }\textbf {\bibinfo {volume}
  {17}},\ \bibinfo {pages} {065009} (\bibinfo {year}
  {2015}{\natexlab{b}})}\BibitemShut {NoStop}%
\bibitem [{\citenamefont {Acevedo}\ \emph {et~al.}(2017)\citenamefont
  {Acevedo}, \citenamefont {Safavi-Naini}, \citenamefont {Schachenmayer},
  \citenamefont {Wall}, \citenamefont {Nandkishore},\ and\ \citenamefont
  {Rey}}]{Acevedo2017}%
  \BibitemOpen
  \bibfield  {author} {\bibinfo {author} {\bibfnamefont {O.~L.}\ \bibnamefont
  {Acevedo}}, \bibinfo {author} {\bibfnamefont {A.}~\bibnamefont
  {Safavi-Naini}}, \bibinfo {author} {\bibfnamefont {J.}~\bibnamefont
  {Schachenmayer}}, \bibinfo {author} {\bibfnamefont {M.~L.}\ \bibnamefont
  {Wall}}, \bibinfo {author} {\bibfnamefont {R.}~\bibnamefont {Nandkishore}},\
  and\ \bibinfo {author} {\bibfnamefont {A.~M.}\ \bibnamefont {Rey}},\ }\href
  {https://doi.org/10.1103/PhysRevA.96.033604} {\bibfield  {journal} {\bibinfo
  {journal} {Phys. Rev. A}\ }\textbf {\bibinfo {volume} {96}},\ \bibinfo
  {pages} {033604} (\bibinfo {year} {2017})}\BibitemShut {NoStop}%
\bibitem [{\citenamefont {Zhu}\ \emph {et~al.}(2019)\citenamefont {Zhu},
  \citenamefont {Rey},\ and\ \citenamefont {Schachenmayer}}]{Zhu2019}%
  \BibitemOpen
  \bibfield  {author} {\bibinfo {author} {\bibfnamefont {B.}~\bibnamefont
  {Zhu}}, \bibinfo {author} {\bibfnamefont {A.~M.}\ \bibnamefont {Rey}},\ and\
  \bibinfo {author} {\bibfnamefont {J.}~\bibnamefont {Schachenmayer}},\ }\href
  {https://doi.org/10.1088/1367-2630/ab354d} {\bibfield  {journal} {\bibinfo
  {journal} {New J. Phys.}\ }\textbf {\bibinfo {volume} {21}},\ \bibinfo
  {pages} {082001} (\bibinfo {year} {2019})}\BibitemShut {NoStop}%
\end{thebibliography}%

\end{document}